\documentclass[%
 aip,
 rsi,
amsmath,amssymb,
reprint,
floatfix
]{revtex4-1}
\usepackage{graphicx}% Include figure files
\usepackage{dcolumn}% Align table columns on decimal point
\usepackage{bm}% bold math
\usepackage[utf8]{inputenc}
\usepackage{mathptmx}
\usepackage{multirow}
\usepackage{braket}
\usepackage[english]{babel}
\usepackage{xcolor}% Include colors for document elements
\colorlet{RED}{red}
\colorlet{BLUE}{blue}
\usepackage{dcolumn}% Align table columns on decimal pointhttps://v1.overleaf.com/17481932tmwjwzytcbcq#
\usepackage{bm}% bold math
\usepackage{acronym}
\usepackage{adjustbox}
\usepackage{float}
\usepackage{tikz}
\usetikzlibrary{quantikz}
\usepackage{microtype} % fixes many of the overflow hbox errors
\usetikzlibrary{calc,shapes.geometric,decorations.pathmorphing,patterns}

\definecolor{background-color}{gray}{0.98}
\usepackage[margin=2.3cm,bmargin=1cm,footnotesep=1cm]{geometry}

\DeclareFontFamily{U}{mathx}{}
\DeclareFontShape{U}{mathx}{m}{n}{<-> mathx10}{}
\DeclareSymbolFont{mathx}{U}{mathx}{m}{n}
\DeclareMathAccent{\widehat}{0}{mathx}{"70}
\DeclareMathAccent{\widecheck}{0}{mathx}{"71}
\DeclareMathAccent{\widebar}{0}{mathx}{"73}

\newcommand{\PNNL}{%
   \affiliation{%
       Physical Sciences Division, Pacific Northwest National Laboratory, Richland, WA 99354, USA
       }}

\begin{document}

\title{Fock-space Schrieffer--Wolff transformation: classically-assisted rank-reduced quantum phase estimation algorithm.}

\author{Karol Kowalski} 
\PNNL

\author{Nicholas P. Bauman} 
\PNNL

\date{\today}

\begin{abstract}
We present an extension of many-body downfolding methods to reduce the resources required in the quantum phase estimation (QPE) algorithm. In this paper, we focus on the Schrieffer--Wolff (SW) transformation of the electronic Hamiltonians for molecular systems that provides significant simplifications of quantum circuits for simulations of quantum dynamics. We demonstrate that  by employing
Fock-space variants of the SW transformation (or rank-reducing similarity transformations (RRST)) one can
significantly increase the locality of the qubit-mapped similarity transformed Hamiltonians. 
The practical utilization of the SW-RRST formalism is associated with a series of approximations discussed in the manuscript. In particular, amplitudes that define RRST can be evaluated using conventional computers and then encoded on quantum computers. 
The SW-RRST QPE quantum algorithms can also be viewed as an extension of the standard state-specific coupled-cluster downfolding methods to provide a robust alternative to the traditional QPE algorithms to identify the ground and excited states for systems with various numbers of electrons using the same Fock-space representations of the downfolded Hamiltonian. %{\color{blue}
The RRST formalism serves as a design principle for developing 
new classes of approximate schemes that reduce the complexity of quantum circuits.
\end{abstract}

\maketitle

%%%%%%%%%%%%%%%%%%%%%%%%%%%%%%%%%%%%%%%%%%

% NOTES:
% equations for B operator are as  complicated as the quantum problem we want to solve. This is only a platfrom for deriving approximations.

\section{Introduction}

The coupled-cluster (CC) theory \cite{coester58_421,coester60_477,cizek66_4256,paldus72_50,purvis82_1910,arponen83_311,bishop1991overview,jorgensen90_3333,paldus07,crawford2000introduction,bartlett_rmp} has assumed a preeminent role in 
providing a high-accuracy description of diversified classes of many-body systems \cite{arponen83_311,arponen1987extended1,arponen1987extended,arponen1991independent,arponen1993independent,robinson1989extended,arponen1991holomorphic,emrich1984electron} 
,
quantum field theory,\cite{funke1987approaching,kummel2001post,hasberg1986coupled,bishop2006towards,ligterink1998coupled} quantum hydrodynamics,\cite{arponen1988towards,bishop1989quantum}
nuclear structure theory,\cite{PhysRevC.69.054320,PhysRevLett.92.132501,PhysRevLett.101.092502} quantum chemistry,\cite{scheiner1987analytic,sinnokrot2002estimates,slipchenko2002singlet,tajti2004heat,crawford2006ab,parkhill2009perfect,riplinger2013efficient,yuwono2020quantum} and material sciences.\cite{stoll1992correlation,hirata2004coupled,katagiri2005equation,booth2013towards,degroote2016polynomial,mcclain2017gaussian,
wang2020excitons,PhysRevX.10.041043,farnell2004coupled,farnell2018interplay,bishop2019frustrated}  
Many  strengths of the single-reference CC formalism (SR-CC)  or coupled-cluster methods originate in the exponential parametrization of the ground-state wave function 
and closely related linked cluster theorem.\cite{brandow67_771,lindgren12}

The standard CC downfolding techniques \cite{safkk,bauman2019downfolding,bauman2019quantumex,downfolding2020t,kowalski2021dimensionality,bauman2022coupled,
bauman2022coupled2c,
he2022second} provide a many-body form of the effective (downfolded) Hamiltonians that can be used to calculate ground-state energies in reduced-dimensionality active spaces as long as the so-called external amplitudes defining the ground-state out-of-active-space  correlation effects are known or can be effectively approximated. 
Although these methods have originated in the context of single-reference CC  theory leading to active-space representations of non-Hermitian effective Hamiltonians, it became clear that the utilization of the double unitary CC (DUCC) Ansatz can provide Hermitian formulations for downfolded/effective active-space Hamiltonians, which thereafter have intensively been tested and validated in the context of quantum simulations based on the utilization of various quantum solvers. For example, the quantum phase estimation (QPE) \cite{luis1996optimum,
cleve1998quantum,berry2007efficient,childs2010relationship,wecker2015progress,
haner2016high,poulin2017fast} and variational quantum eigensolvers (VQE) \cite{peruzzo2014variational,mcclean2016theory,romero2018strategies,PhysRevA.95.020501,Kandala2017,kandala2018extending,PhysRevX.8.011021,huggins2020non,ryabinkin2018qubit,cao2019quantum,ryabinkin2020iterative,izmaylov2019unitary,lang2020unitary,grimsley2019adaptive,grimsley2019trotterized,cerezo2021variational,mcardle2020quantum,bharti2022noisy} were invoked to obtain ground-state energies of molecular systems. 
These tests demonstrated that DUCC-based downfolded Hamiltonians and corresponding dimensionality reduction could accurately reproduce the electronic energies for basis sets of the sizes that are currently beyond the reach of the most advanced quantum algorithms and quantum hardware.\cite{kivlichan2018quantum,low_depth_Chan,google2020hartree,mcardle2020quantum,bharti2022noisy} 
The Hermitian CC downfolding procedures were also discussed and tested in the context of quantum dynamics and excited-state simulations.\cite{bauman2019quantumex,downfolding2020t,bauman2022coupled} Due to the state-specificity of the downfolding procedures, the latter attempts require detailed knowledge of the external Fermionic degrees of freedom for excited states and the construction of separate effective Hamiltonians for each excited state. Although extraction of excited-state external correlation effects is possible for some classes of excited states, which can be captured by approximate Equation-of-Motion CC (EOMCC) methods,\cite{bartlett89_57,bartlett93_414,stanton93_5178}
the generalization of this formalism to general-type excited states may be numerically challenging. 
% OKKK
A part of the problem is also associated with translating the EOMCC Ansatz defined by commuting operators into a language of unitary CC expansions involving non-commuting operators. 
% OKKK

Instead of following this strategy, in this paper, we discuss the class of Fock-space Schrieffer--Wolff (SW) transformation-inspired downfolding procedures designed to simplify the many-body form of the Hamiltonian. 
In analogy to the standard DUCC-based  techniques, the  SW-transformation-based 
(or  rank-reducing similarity transformation (RRST))  formulations utilize the partitioning of one-electron states (spin-orbitals) into active and external spin-orbitals. Although the RRST cannot eliminate  all components of the Hamiltonian that involve creation/annihilation operators carrying the external spin-orbital indices (the so-called external component of the Hamiltonian), the RRST is designed in a way that leads to a simple form that involves only local actions of qubits in 
corresponding quantum algorithms such as QPE. 
% OKKK
In the context of QPE methodology, this form not only makes qubit mappings simpler but can enable more efficient utilization of the Trotter formulas. In contrast to the standard DUCC downfolding, the proposed approach and related approximations eliminate its state-specific character and provide a description of multiple electronic states corresponding to the ground and excited states at least well approximated by the set of active orbitals. An exciting feature of the discussed framework is its universal character (in the sense of Fock space) in describing many-body systems with various numbers of electrons (particles), where the number of particles is specified when the action of the Hamiltonian (or the corresponding quantum evolution operator) on specific states takes place. The discussed development is primarily motivated by impressive progress in developing Fock-space generalization of CC formulations.\cite{jeziorski1989valence,meissner1996multiple,meissner1998fock,musial2008intermediate,meissner2022new}

In analogy to all existing algorithms, the RRST formalism can be viewed as a platform for developing broad classes of approximations. In this paper, we will outline the hybrid algorithm that combines the classical-computing part associated with the determination of the RRST and the quantum-computing part, which provide a mean for modeling time evolution generated by the RRST downfolded Hamiltonian. 
%We will also focus on alleviating possible numerical problems in classical parts by taking advantage of the flexibility in defining similarity transformation  - for example, by introducing  second similarity transformation (which can be viewed as a gauge transformation),  and imposing certain conditions on its form. 
% OKKK

%%%%%%%%%%%%%%%%%%%%%%%%%%%%%%%%%%%%%%%%%%
\section{Rank-reducing Unitary Similarity Transformations of many-body Hamiltonians}

The dynamics of the quantum system are given by the evolution operator $\Omega(t)$ 
\begin{equation}
   \Omega(t)= e^{-itH} 
    \label{eq1}
\end{equation}
where we assume that the Hamiltonian $H$ is time-independent
and takes the following second-quantized form in the 
basis of $N$ spin-orbitals 
% OKKK
\begin{equation}
H=\sum_{p,q=1}^N h^p_q a_p^{\dagger}a_q + \frac{1}{4} \sum_{p,q,r,s=1}^N
v^{pq}_{rs} a_p^{\dagger} a_q^{\dagger} a_s a_r \;,
\label{eq2}
\end{equation}
where $p,q,r,s$ are spin-orbital indices and $h^p_q$ and $v^{pq}_{rs}$ are one- and two-electron (anti-symmetrized) integrals defining $H_1$ and $H_2$ operators, respectively. The $a_p^{\dagger}$ ($a_p$) operator corresponds to a creation (annihilation) operator for the electron in $p$-th spin-orbital.
In quantum computing applications, especially in quantum phase estimations, various representations of the electronic Hamiltonian (induced, for example,  by the unitary transformations) can be used. 
This is a consequence of the fact that the spectrum of the Hamiltonian remains unchanged upon these transformations.  Let us denote a general unitary transformation $U$ as
% OKKK
\begin{equation}
    U=e^{A(t)} \;,
    \label{eq3}
\end{equation}
where the operator $A(t)$ is anti-Hermitian
\begin{equation}
    A(t)^{\dagger} = -A(t) \;,
    \label{eq4}
\end{equation}
and similarity-transformed Hamiltonian $\bar{H}(t)$ is defined as 
\begin{equation}
     \bar{H}(t)=e^{-A(t)}H(t)e^{A(t)} \;.
    \label{eq5}
\end{equation}
In analogy to the $H$ operator,  the similarity-transformed Hamiltonian is also Hermitian. 
In consequence, probing the phase with the  $\bar{H}$ and corresponding 
time evolution operator $\bar{\Omega}(t)$
\begin{equation}
    \bar{\Omega}(t)=e^{-it \bar{H}(t)}=e^{-A(t)} \Omega(t) e^{A(t)} 
    \label{eq6}
\end{equation}
should detect the same values of energy/phase (subject to the various choices of initial state). A typical illustration of the above techniques is the interaction and Heisenberg pictures widely used in quantum mechanics, which have recently been explored in the context of quantum computing simulations. The interaction-picture-based approach has recently been studied in the context of quantum computing.\cite{low2018hamiltonian,rajput2021hybridized,watkins2022time}
% OKKK

\begin{figure}
\includegraphics[width=10.0 cm]{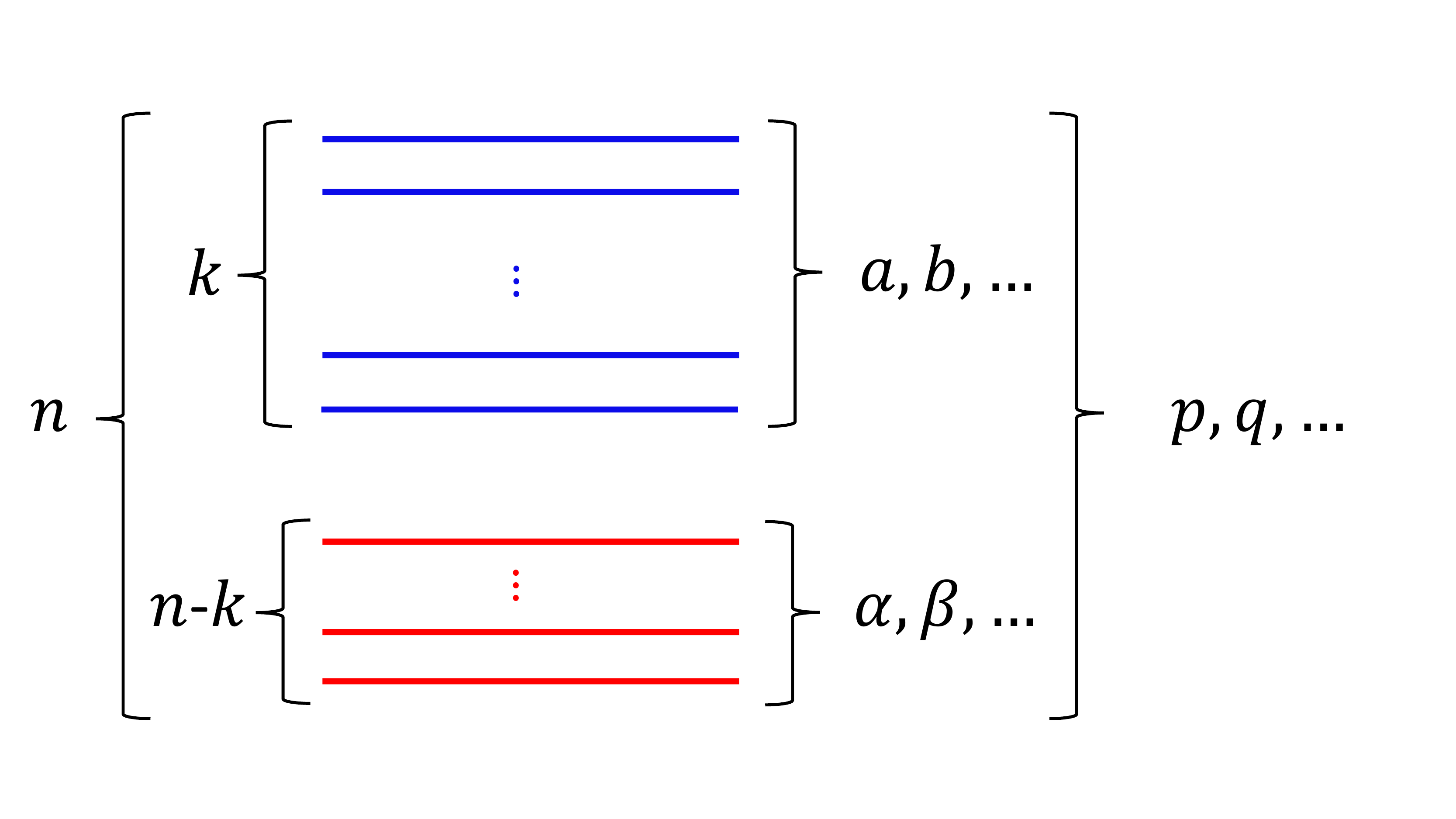}
\caption{The orbital (spin-orbital)  domain is partitioned into $n-k$ active and $k$ external orbitals ($n$ stands for the total number of orbitals). The active spin-orbitals are denoted as $\alpha$, $\beta$, ...,
the external spin-orbitals as $a$, $b$, ..., and generic spin-orbitals as 
$p$, $q$, ... . In general, the active spin-orbitals do not have to be defined as the "lowest" lying spin-orbitals using some energy-related criteria. At this moment, the nature of active/external spin-orbitals remains unspecified. The total number of spin-orbitals $N$ is defined as $N=2n$.}
\label{fig1}
\end{figure}

\begin{figure}
\includegraphics[width=11.0 cm]{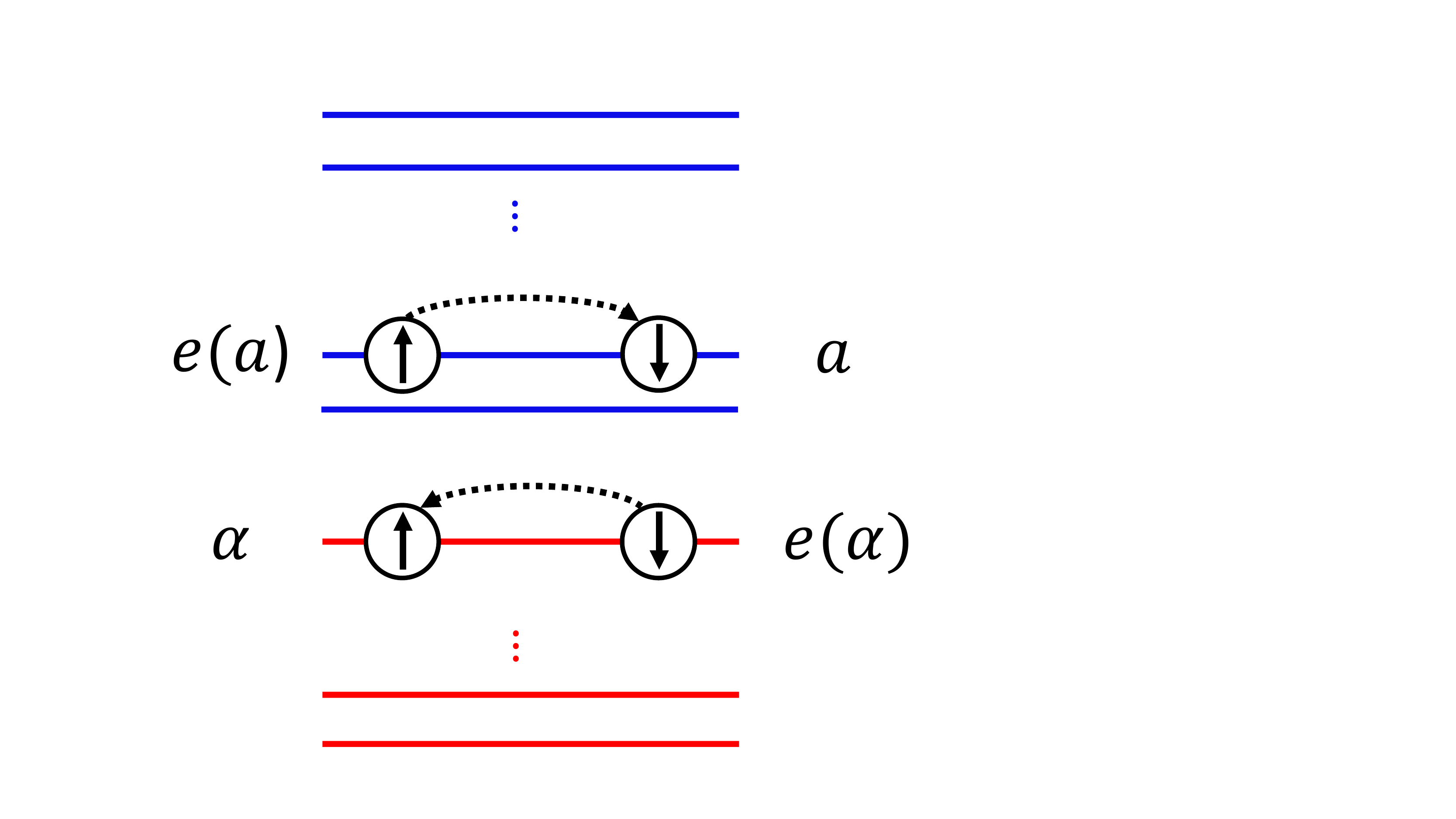}
\caption{A typical example of the external off-diagonal isoenergetic excitation operator 
$E^{\alpha a}_{e(\alpha) e(a)}$.}
\label{fig2}
\end{figure}

In this paper, we will pursue a slightly different goal associated with the design of  {\it time-independent} unitary transformation generated by the time-independent anti-Hermitian operator $B$  ($B^\dagger = -B$) such that 
\begin{equation}
    H=e^{-B} G e^{B}\;,
    \label{eq6a}
\end{equation}
where the properties of the $G$ operator and the form of the $B$ operator assuring these properties will be discussed later.  For the sake of the following discussion, let us introduce the partitioning of the orbitals (spin-orbitals) into active (with corresponding first  $2(n-k)$ qubits) and external
(with qubits enumerated as $2(n-k)+1,\ldots,2n$)
as shown in Fig.\ref{fig1}.
% OKKK
Additionally, we will assume that all spin-orbitals are arranged in a way that spin-up ($\uparrow$) and spin-down ($\downarrow$) spin-orbitals occupying the same orbitals $P$ (isoenergetic spin-orbitals) are neighboring as shown in the following scheme
\begin{equation}
    \ldots [Q\uparrow][Q\downarrow] \ldots  [P\uparrow][P\downarrow] \ldots
    \label{eq7}
\end{equation}
If spin-orbitals $p$ and $q$ are isoenergetic we will denote it by 
$q=e(p)$ (or $p=e(q)$).
% OKKK
%
%
%
\unskip
%
%
%
% active space in analogy to Fock-space CC formulations can be chosen to provide 0-th order  description of several systems with different numbers of active electrons such as in the valence universal Fock-space formulations
%
%
%
The two classes of spin-orbitals induce the partitioning of the second-quantized operators into internal and external parts that are defined by   creation-annihilation operator strings
(CAOSs)
carrying only active spin-orbital indices (active part) and strings that contain at least one creation/annihilation operator carrying external spin-orbital index (external part), respectively.  The internal and external 
parts of arbitrary operator $X$ can be symbolically denoted as 
\begin{equation}
    \widecheck{\mathcal{P}}(X)
    \label{eq8}
\end{equation}
for the internal part and 
\begin{equation}
    \widehat{\mathcal{P}}(X)
    \label{eq9}
\end{equation}
for the external part. Typical examples of CAOSs  entering internal and external parts are $E^{\alpha}_{\beta}=a^{\dagger}_{\alpha} a_{\beta}$
and $E^{\alpha}_{b}=a^{\dagger}_{\alpha} a_{b}$, respectively, where the general form of the excitation operator $E^{pq\ldots}_{rs\ldots}$ is defined as   $E^{pq\ldots}_{rs\ldots}=a_p^{\dagger} a_q^{\dagger} \ldots a_s a_r$ (the $E^{pq\ldots}_{rs\ldots}$ operator is antisymmetric with respect to swapping adjacent spin-orbital indices). Furthermore, it is convenient to decompose the 
external part into its diagonal part ($\widehat{\mathcal{P}}_d(X)$), iso-energetic  off-diagonal
% modity isoenergetic part also here
($\widehat{\mathcal{P}}_{ie}(X)$), and energetically distinct off-diagonal
($\widehat{\mathcal{P}}_{eod}(X)$)
that is defined by the following classes of CAOSs (in all $E^{pq\ldots}_{rs\ldots}$ below, we assume that $p<q<\ldots$ and $r<q<\ldots$):
%% Change the definition of P_{id} - off-diagonal that contain isoenergietic excitations
\begin{widetext}
\begin{eqnarray}
   \widehat{\mathcal{P}}_d(X) &\rightarrow& \lbrace  E^a_a, E^{\alpha a}_{\alpha a}, E^{ab}_{ab}, \ldots \rbrace  \;, \label{eq10} \\
   \widehat{\mathcal{P}}_{ie}(X) &\rightarrow& \lbrace E^a_{e(a)}, 
   E^{\alpha a}_{e(\alpha) e(a)}, E^{ab}_{e(a)e(b)}, \ldots   \rbrace \;, \label{eq11} \\
   \widehat{\mathcal{P}}_{eod}(X)  &\rightarrow&   
   \lbrace E^a_b, E^{ac}_{bd}, \ldots \rbrace \;\; (\mbox{all off-diagonal CAOSs that are not isoenergetic})\;.\label{eq12}
\end{eqnarray}
\end{widetext}
%It is also assumed that the operator classes defining $\widehat{\mathcal{P}}_d(X)$, $\widehat{\mathcal{P}}_{ie}(X)$, and  $\widehat{\mathcal{P}}_{eod}(X)$ sets have to reflect all symmetries (spin and spatial) of the $X$ operator, therefore definitions (\ref{eq10})-(\ref{eq12}) are of general nature. 
% WORK ON THIS PART
%{\color{blue}
The isoenergetic part, $\widehat{\mathcal{P}}_{ie}(X)$, is defined by off-diagonal external excitation operators where 
the sums of energies corresponding to all upper spin-orbitals and energies corresponding to lower spin-orbital indices are equal according to the spin-orbital energy ordering shown in Fig.\ref{fig1}.
For example, this includes situations when
each upper spin-orbital index ($p$) has a corresponding isoenergetic lower spin-orbital index ($e(p)$) according to the spin-orbital energy ordering shown in Fig.\ref{fig2}.
%}
%{\color{blue}The isoenergetic part, $\widehat{\mathcal{P}}_{ie}(X)$, is defined by off-diagonal external excitation operators where each upper spin-orbital index has a corresponding isoenergetic lower spin-orbital index according to the spin-orbital energy ordering shown in Fig.\ref{fig2}.}
% $E^{pq\ldots}_{rs\ldots}$ where the sums of energies corresponding to all upper spin-orbitals and energies corresponding to lower spin-orbital indices are equal according to the spin-orbital energy ordering shown in Fig.\ref{fig1}. 
%For characterizing norms of the Hamiltonian in various operator spaces, 
We will also define external off-diagonal part, $\widehat{\mathcal{P}}_{od}(X)$, defined as 
\begin{equation}
    \widehat{\mathcal{P}}_{od}(X)=\widehat{\mathcal{P}}_{ie}(X)+
    \widehat{\mathcal{P}}_{eod}(X) \;.
\label{eq12xx}
\end{equation}
For example, using these decompositions, the electronic Hamiltonian $H$ 
(for simplicity, in this manuscript, we focus on the spin-independent Hamiltonians)
can be decomposed as 
% OKKK
\begin{widetext}
\begin{equation}
    H=\widecheck{\mathcal{P}}(H)+\widehat{\mathcal{P}}(H)=\widecheck{\mathcal{P}}(H)+\widehat{\mathcal{P}}_d(H)+\widehat{\mathcal{P}}_{ie}(H)+\widehat{\mathcal{P}}_{eod}(H) \label{eq13}
\end{equation}
\end{widetext}
where 
\begin{widetext}
\begin{eqnarray}
   \widecheck{\mathcal{P}}(H) &=& \sum_{\alpha\beta} h^{\alpha}_{\beta} E^{\alpha}_{\beta} +
   \sum_{\alpha<\beta;\gamma<\delta} v^{\alpha\beta}_{\gamma\delta} E^{\alpha\beta}_{\gamma\delta}   \label{eq14} \\
   \widehat{\mathcal{P}}_d(H) &=& \sum_a h^a_a E^a_a  + \sum_{\alpha a} v^{\alpha a}_{\alpha a} E^{\alpha a}_{\alpha a} 
   +\sum_{a<b} v^{ab}_{ab} E^{ab}_{ab}  \label{eq15} \\
   \widehat{\mathcal{P}}_{ie}(H) &=& 
   \sum_{a} h^{a}_{e(a)} E^a_{e(a)} +
   \sum_{\alpha a} v^{\alpha a}_{\alpha a} E^{\alpha a}_{e(\alpha) e(a)}
   + \sum_{a<b} v^{ab}_{ab} E^{ab}_{e(a) e(b)} \label{eq16} \\
   \widehat{\mathcal{P}}_{eod}(H) &=& \widebar{\sum_{\alpha a}} h^{\alpha}_{a}E^{\alpha}_{a}+
   \widebar{\sum_{a\alpha}} h^{a}_{\alpha}E^{a}_{\alpha}  
   +\widebar{\sum_{a b}} h^{a}_{b}E^{a}_{b}+
   \widebar{\sum_{\alpha b;\gamma<\delta}} v^{\alpha b}_{\gamma\delta} E^{\alpha b}_{\gamma\delta}  \nonumber \\
   && +\widebar{\sum_{\alpha<\beta;\gamma d}} v^{\alpha\beta}_{\gamma d} E^{\alpha\beta}_{\gamma d}+
   \widebar{\sum_{\alpha<\beta;c<d}} v^{\alpha \beta}_{cd} E^{\alpha\beta}_{cd}  
   \widebar{\sum_{a<b;\gamma<\delta}} v^{ab}_{\gamma\delta} E^{ab}_{\gamma\delta} 
    + \widebar{\sum_{\alpha b;\gamma d}} v^{\alpha b}_{\gamma d} E^{\alpha b}_{\gamma d} \nonumber \\
   &&+
   \widebar{\sum_{a<b;\gamma d}} V^{ab}_{\gamma d} E^{ab}_{\gamma d} +
   \widebar{\sum_{\alpha b;c<d}} V^{\alpha a}_{cd}  E^{\alpha a}_{cd} 
   + \widebar{\sum_{a<b;c<d}} V^{ab}_{cd}  E^{ab}_{cd} \label{eq17}
\end{eqnarray}
\end{widetext}
and $\widebar{\sum}$ symbol represents the summation over off-diagonal non-isoenergetic terms. 
Assuming that $n-k < k$,
one can see that $\widehat{\mathcal{P}}_{eod}(H)$ includes the largest number of terms of all components defining the decomposition (\ref{eq13}). It is also obvious that, in general, the $\widehat{\mathcal{P}}_d(X)$ of an arbitrary operator $X$ 
can be expressed in terms of particle number operators, $n_p$,
\begin{equation}
 n_p = a_p^{\dagger} a_p \;,
 \label{eq18}
\end{equation}
which can symbolically be denoted as 
\begin{equation}
    \widehat{\mathcal{P}}_d(X) = f_X(\lbrace n_p \rbrace_{p=1}^{N}).
    \label{eq19}
\end{equation}
For example, 
\begin{eqnarray}
    \widehat{\mathcal{P}}_d(H) &=& f_H(\lbrace n_p \rbrace_{p=1}^{N})  \nonumber \\
    &=&
    \sum_a h^a_a n_a  + \sum_{\alpha a} v^{\alpha a}_{\alpha a} n_{\alpha} n_a +
    \sum_{a b} v^{ab}_{ab} n_{a} n_b \;.
    \label{eq20}
\end{eqnarray}
\unskip
% OKKK
%
%OKKK

\section{Schrieffer--Wolff rank-reducing similarity transformations}

In this Section, we discuss the possibility of designing unitary transformation generated by the time-independent 
anti-Hermitian operator $B$ (see Eq.(\ref{eq6a})
\begin{equation}
    G=e^{B} H e^{-B}\;,
    \label{eq6ab}
\end{equation}
that assures specific properties of the $G$ operator and consequently simplifies qubit mapping of the evolution operator $\Omega(t)$. 
% OKKK
%%% makes it analog !!!!!!!!!!!!!!! 0/1 flips 
%%%% k-orbitals or 2k-qubits can be more "classical"
%%% two types of qubit
%%% Quantum MPI (ideal)
In particular, we will prove the following Theorem:\\
%%\begin{Theorem}
{\bf Theorem 1.} {\em There exist conditions for anti-Hermitian $B$-operator that render the $G$ operator in the following form: $G=\widecheck{\mathcal{P}}(G)+\widehat{\mathcal{P}}_d(G)+\widehat{\mathcal{P}}_{ie}(G)$.
Moreover, the action of $\widehat{\mathcal{P}}_d(G)+\widehat{\mathcal{P}}_{ie}(G)$ on the qubits corresponding to external spin-orbitals (assuming the ordering of spin-orbitals and qubits in accordance with Eq.(\ref{eq7}))
is local.} \\
%%\end{Theorem}
% OKKK
%
%\begin{proof}[Proof of Theorem 1]
{\bf Proof of Theorem 1}
To eliminate the $\widehat{\mathcal{P}}_{eod}(G)$ we impose the condition for the $B$ operator
\begin{equation}
    \widehat{\mathcal{P}}_{eod}(G)=\widehat{\mathcal{P}}_{eod}(e^BHe^{-B}) = 0 \;,
    \label{eq22}
\end{equation}
%{\color{red} the dimension of space where you solving/approximating these equations is equivalent to the total terms in  $\widehat{\mathcal{P}}_{eod}$ space. Comment on this - perturbative analysis is very helpful. Solving equations in Fock space should be carefully designed ....} 
which can explicitly be expanded in terms of multi-commutator expansion using Hausdorff expansion 
\begin{equation}
   \widehat{\mathcal{P}}_{eod}(H) - 
   \widehat{\mathcal{P}}_{eod}([H,B])
   +\frac{1}{2} \widehat{\mathcal{P}}_{eod}([[H,B],B])
   - \ldots = 0. 
   \label{eq23}
\end{equation}
Solving the above equations in the Fock space requires some attention. One should realize that the dimensionality of  equations (\ref{eq23}) is much bigger than the number of non-zero terms contributing to 
$\widehat{\mathcal{P}}_{eod}(H)$ in Eq.(\ref{eq23}) and is equal to the number of all excitation-type operators spanning 
$\widehat{\mathcal{P}}_{eod}$ space in Eq.(\ref{eq12}).
The perturbative analysis of the solution of Eq.(\ref{eq22}) (see Appendix A) shows that at the first order of perturbation theory,   the number of amplitudes defining $B$ is precisely equal to the number of non-zero elements of $\widehat{\mathcal{P}}_{eod}(H)$. 
However, higher orders of many-body perturbation theory generate higher many-body components in space $\widehat{\mathcal{P}}_{eod}$.
For this reason, in the general case, the $B$ operator is expressed in terms of all operators in set (\ref{eq12})
and satisfy the condition
%The minimal form of the $B$ operator must assure that the number of equations in equal to the number of equations, equal to the number of terms in $\widehat{\mathcal{P}}_{eod}(H)$, which is satisfied when 
\begin{equation}
    B=\widehat{\mathcal{P}}_{eod}(B) \;,
    \label{24}
\end{equation}
which guarantees that the number of equations and unknowns are equal. As will be discussed later, various approximate techniques can be used to approximate the $B$ operator using a smaller number of variables. 
If the solution of  Eq.(\ref{eq22}), labeled as $B^{\star}$, can be found (or effectively approximated by limiting the rank of the multi-commutator expansion) on classical computers, then 
the $G$ operator with the desired properties is given by
\begin{equation}
G=\widecheck{\mathcal{P}}(e^{B^{\star}} H e^{-B^{\star}})+\widehat{\mathcal{P}}_d(e^{B^{\star}} H e^{-B^{\star}})+\widehat{\mathcal{P}}_{ie}(e^{B^{\star}} H e^{-B^{\star}})\;.
\label{eq25}
\end{equation}
The  parts of $G$ contributing to action on the qubits corresponding to external spin-orbitals are 
$\widehat{\mathcal{P}}_d(e^{B^{\star}} H e^{-B^{\star}})$ and $\widehat{\mathcal{P}}_{ie}(e^{B^{\star}} H e^{-B^{\star}})$. %{\color{blue} 
These two classes of elements are advantageous components of $G$ when designing quantum circuits with low complexity.
%} 
The diagonal part, according to Eq.(\ref{eq19}), obviously involves the particle number operators that are qubit-local.
On the other hand, the encoding of general operators %(\ref{eq21}) 
contributing to the 
$\widehat{\mathcal{P}}_{ie}(e^{B^{\star}} H e^{-B^{\star}})$ requires encoding a chain of 
$E^{a}_{e(a)}$/$E^{\alpha}_{e(\alpha)}$ and their Hermitian conjugates. It turns out, however, that 
these operators involve gates only on the pairs of adjacent qubits 
$[Q\uparrow][Q\downarrow]$
in the representation given by scheme (\ref{eq7}).
It can be easily inspected by using for example Jordan--Wigner (JW)  qubit encoding \cite{Jordanwigner1928}
\begin{eqnarray}
  a_p^{\dagger} &\rightarrow & Q_p^{+} \otimes Z^{\rightarrow}_{p-1} \;,\label{jw1} \\
  a_p &\rightarrow & Q_p^{-} \otimes Z^{\rightarrow}_{p-1} \;,\label{jw}
\end{eqnarray}
where 
\begin{eqnarray}
   Q^{+} &=& \frac{1}{2}(\sigma_p^x-i\sigma_p^y) \;, \label{jw3} \\
   Q^{-} &=& \frac{1}{2}(\sigma_p^x+i\sigma_p^y) \;, \label{jw4} \\
   Z^{\rightarrow}_{p-1} &=& \sigma^z_{p-1}\otimes \ldots \otimes \sigma^z_1 \;, \label{jw5}
\end{eqnarray}
and $\sigma^x_p$, $\sigma^y_p$, and $\sigma^z_p$ represent Pauli gates on $p$-th qubit, that the general  $a^{\dagger}_{Q\uparrow}a_{Q\downarrow}$ can be expressed as local two-qubit action
\begin{equation}
    I_{Q\uparrow+1}^{\leftarrow}\otimes Q^+_{Q\uparrow}\otimes Q^-_{Q\downarrow}\otimes I_{Q\downarrow-1}^{\rightarrow}  \;.
    \label{jw6}
\end{equation}
where $I_{Q\uparrow+1}^{\leftarrow}$ and $I_{Q\downarrow-1}^{\rightarrow}$ symbolically represent tensor products of the unit operator to the left and to the right of the $Q\uparrow$ and $Q\downarrow$ qubits, respectively.
$\square$
%\end{proof}

An interesting consequence of Eq.(\ref{eq22}) is the fact that the solution $B^{\star}$ cannot commute with the Hamiltonian $H$.  If it was the case, i.e.,
\begin{equation}
    [B^{\star},H]=[e^{B^{\star}},H]=0
    \label{add1}
\end{equation}
then the corresponding equation (Eq.(\ref{eq22})), which becomes 
\begin{equation}
   \widehat{\mathcal{P}}_{eod}(H)=0 \;,
    \label{add2}
\end{equation}
has no solutions (where we assumed the non-trivial case of a Hamiltonian
$\widehat{\mathcal{P}}_{eod}(H)\ne 0$).

The unitary transformation generated by the $B$ operator was used to eliminate the most non-local, in the sense of qubit utilization, $\widehat{\mathcal{P}}_{eod}(G)$  part of the $G$ operator. 
%{\color{blue}
In a similar fashion, one can also eliminate the $\widehat{\mathcal{P}}_{ie}(G)$ contribution to the external excitations. However, an added layer of complexity is associated with the numerical nature of the problem that must be considered. For example, certain classes of solvers, like many-body perturbations theory, may stumble into numerical problems associated with the vanishing denominators for isoenergetic components of $B$. These problems can be remedied (in addition to using zeroth order Hamiltonians that break symmetries of the system) by using  additional unitary "gauge" transformation and breaking the energetic symmetry of the external isoenergetic off-diagonal terms ({\it vide infra}).  If the numerical issues can be effectively handled, then  the "domain" of the $B$ operator can be extended to the external isoenergetic off-diagonal excitations, and the following Corollary holds: \\
%}
%
%%%%%%%%%%%%%%%%%%%%%%\begin{Corollary}
{\bf Corollary.} {\em If equations   can be solved for the $B$ operator with the solution $B^{\star}$ in the extended excitations  domain involving isoenergetic external off-diagonal excitations, i.e., 
\begin{equation}
 \widehat{\mathcal{P}}_{ie}(e^BHe^{-B}) + \widehat{\mathcal{P}}_{eod}(e^BHe^{-B}) = 
 \widehat{\mathcal{P}}_{od}(e^BHe^{-B})=
 0
 \label{eq26}
\end{equation}
then the 
operator $G$ takes the form 
\begin{equation}
G=\widecheck{\mathcal{P}}(e^{B^{\star}} H e^{-B^{\star}})+\widehat{\mathcal{P}}_d(e^{B^{\star}} H e^{-B^{\star}})\;,
\label{eq27}
\end{equation}
where are $\widehat{\mathcal{P}}_d(e^{B^{\star}} H e^{-B^{\star}})$ is expressed solely  in terms of the particle number operators, i.e., \linebreak
$\widehat{\mathcal{P}}_d(e^{B^{\star}} H e^{-B^{\star}})=f_G(\lbrace n_p \rbrace_{p=1}^{N})$. 
}
%%%%%%%%%%%%%%%%%%%%. \end{Corollary}

So far, we have been discussing the application of the single unitary transformation generated by the unitary operator $e^B$ (or $e^{-B}$ in the context of Eq.(\ref{eq6ab})). However, there exists flexibility in choosing the form of the unitary transformation. For example, the product of two unitary operations, e.g., 
\begin{equation}
U=e^B e^C \;,
\label{eq28}
\end{equation}
%{\color{red} Check U " e(C)e(B) or e(B)e(C)}
where $C^{\dagger}=-C$,
can transform $H$ operator to $\Gamma$ operator in analogous was in  Eq.(\ref{eq6ab}),
\begin{equation}
    \Gamma=e^B e^C H e^{-C} e^{-B}
    \label{eq29}
\end{equation}
with the same spectral properties as the original Hamiltonian $H$. However, the purpose of the additional transformation generated by the anti-Hermitian operator $C$ is to produce the form of the auxiliary Hamiltonian $\bar{H}_C$, 
%{\color{red} CHECK C vs -C in exponents conflict between Eq.(40) and (41) in C sign convention}
\begin{equation}
    \bar{H}_C = e^{C}He^{-C} \;,
    \label{eq30}
\end{equation}
that eases the process of solving analogs of the Eq.(\ref{eq22}) 
\begin{equation}
\widehat{\mathcal{P}}_{eod}(e^B\bar{H}_Ce^{-B}) = 0 \;,
\label{eq31}
\end{equation}
or Eq.(\ref{eq26})
\begin{equation}
 \widehat{\mathcal{P}}_{ie}(e^B\bar{H}_Ce^{-B}) + \widehat{\mathcal{P}}_{eod}(e^B\bar{H}_Ce^{-B}) = 
 \widehat{\mathcal{P}}_{od}(e^B\bar{H}_Ce^{-B})
 =
 0 \;.
 \label{eq32}
\end{equation}
For this reason,  we can view  the $e^C$ operator as an auxiliary  transformation  of the Hamiltonian $H$.
The idea behind using the auxiliary transformation is to employ as robust (or even postulated) form of the $C$ operator as possible. In particular, the $C$ operator can include different types of excitation operators than the $B$ operator. For example, it can contain many-body effects in the active space only. 
The utilization of the auxiliary transformation offers us flexibility in exploring various scenarios, including the possibility of spatial/spin symmetry breaking in the simulations of molecular systems, without altering the spectral properties of the original Hamiltonian $H$.

\section{QPE formulations based on the SW-RRST representation of many-body Hamiltonians}

\begin{figure}
\includegraphics[width=6.0 cm]{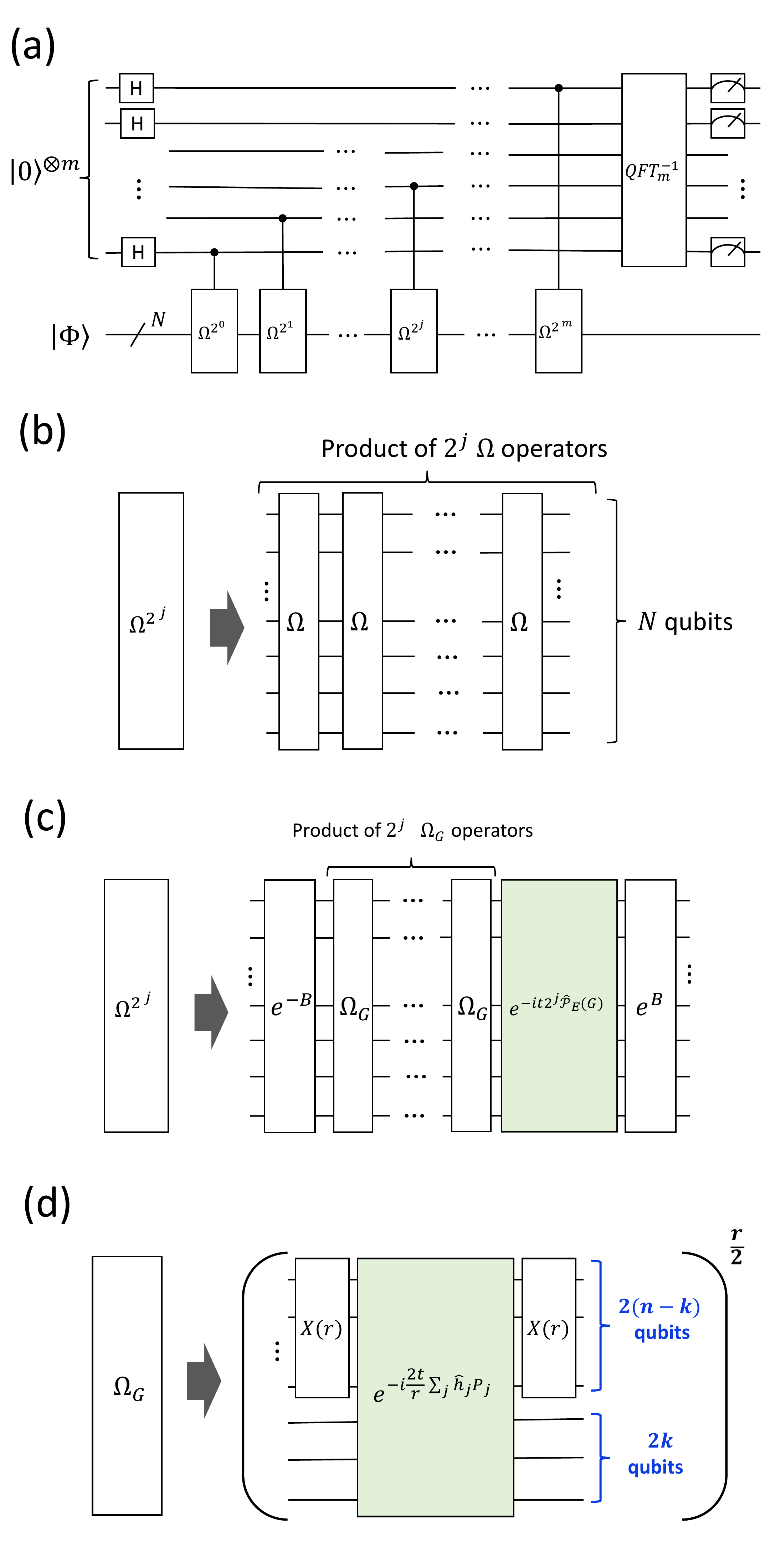}
\caption{Schematic representations of the QPE algorithm and algorithm for the evaluation of the  $2j$-th power of the $\Omega(t)$ operator in the representation involving $G$ operator given by Eq.(\ref{eq6a}). 
%{\color{blue}
See the text for descriptions of each diagram.
%}
}
\label{fig5}
\end{figure}

The main idea of the QPE is in the controlled execution of the powers of the $\Omega(t)$ operator according to the progression 
\begin{equation}
\Omega(t)^{2^0}\rightarrow \Omega(t)^{2^1} \rightarrow 
\ldots \Omega(t)^{2^j} \ldots \ldots \Omega(t)^{2^m}
\label{eq122}
\end{equation}
where $m$ designates the number of ancilla qubits used to read the phase(s) of the unitary evolution operator (see Fig.\ref{fig5}(a)).
When a standard representation of the Hamiltonian $H$ is used, the qubit encoding of  
the $2^j$-th power of $\Omega(t)$ operator,
\begin{equation}
    \Omega(t)^{2^j}= (e^{-itH})^{2^j}\;,
    \label{eq123}
\end{equation}
utilizes all $N$ "physical" and $m$ ancilla qubits. 
When representation (\ref{eq6a}) is invoked, the same operator power can be expressed as 
\begin{equation}
   \Omega(t)^{2^j} = (e^{-B} e^{-itG} e^B )^{2j}=e^{-B} (e^{-itG})^{2^j} e^B \;,
   \label{eq124}
\end{equation}
where $B$ and $G$ operators are given by equations (\ref{eq6a}) or (\ref{eq6ab}). The main difference between (\ref{eq123}) and (\ref{eq124}) (see Fig.\ref{fig5} instes (b) and (c)) is that
the sequence $(e^{-itG})^{2^j}$ can be executed using localized qubit gates in the sense of earlier discussion of $G$-operator properties. The additional advantages stem from the fact that the external part of the $G$ operator 
$\widehat{\mathcal{P}}(G)$ involves a simpler form of the gates and large classes of operators that commute with the internal part of the $G$ operator, $\widecheck{\mathcal{P}}(G)$, which simplifies  the form of the Trotter formula. We will explain it on the example of the $G$ operator discussed in Corollary 1, where $G$ is decomposed into internal $\widecheck{\mathcal{P}}(G)$ and  external $\widehat{\mathcal{P}}(G)$ parts, and where the external part is expressed in terms of number operators 
$\widehat{\mathcal{P}}(G)=f_G(\lbrace n_p \rbrace_{p=1}^{N})$ %{\color{blue} 
since $\widehat{\mathcal{P}}(G)$ only contains diagonal elements as a result of Corollary 1.
%}
Let us further decompose $\widehat{\mathcal{P}}(G)$ into the part that mixes particle number operators for  active and external spin-orbitals ($\widehat{\mathcal{P}}_M(G)$) and the part that is solely expressed in terms of the particle number operator corresponding to external spin-orbitals only ($\widehat{\mathcal{P}}_E(G)$). Therefore the  following commutation 
relations holds 
\begin{equation}
 [\widecheck{\mathcal{P}}(G),\widehat{\mathcal{P}}_E(G)]=
 [\widehat{\mathcal{P}}_M(G),\widehat{\mathcal{P}}_E(G)]=0 \;,
 \label{eq125}
\end{equation}
therefore 
\begin{eqnarray}
    e^{-itG} &=& e^{-it \lbrack \widecheck{\mathcal{P}}(G)+\widehat{\mathcal{P}}_M(G) \rbrack}  e^{-it \widehat{\mathcal{P}}_E(G)} \nonumber \\
    &=& e^{-it \widehat{\mathcal{P}}_E(G)} e^{-it \lbrack \widecheck{\mathcal{P}}(G)+\widehat{\mathcal{P}}_M(G) \rbrack} \;,
    \label{eq125b}
\end{eqnarray}
where the  $e^{-it \lbrack \widecheck{\mathcal{P}}(G)+\widehat{\mathcal{P}}_M(G) \rbrack}$
term requires a Trotter formula to be implemented. At the same time the 
$e^{-it \widehat{\mathcal{P}}_E(G)}$ terms as depending only on the particle number operators (or their products) can be calculated  exactly, which is a consequence of the fact that all $n_p$'s operators defining $\widehat{\mathcal{P}}_E(G)$ commute. For example, 
the JW qubit  mapping of $n_p$ and $n_pn_q$ can be represented by a simple local circuits
\begin{eqnarray}
 n_p &\rightarrow& \frac{1}{2}(1_p-\sigma_p^z) \;, \label{eq126} \\
 n_p n_q &\rightarrow& \frac{1}{4}(1_p-\sigma_p^z)\otimes (1_q-\sigma_q^z) \;,
 \label{eq126b}
\end{eqnarray}
where $1_p$ and  $\sigma_p^z$ are identity and Pauli Z matrices (gates) on $p$-th qubit. We also used a simplified notation, where 
\begin{widetext}
\begin{equation}
    (1_p-\sigma_p^z)\otimes (1_q-\sigma_q^z) \equiv \ldots \otimes 1_i \otimes
     \ldots \otimes (1_p-\sigma_p^z) \otimes\ldots  \otimes 1_k \otimes \ldots \otimes 
     (1_q-\sigma_q^z) \otimes \ldots \otimes  1_l \otimes \ldots \;.
     \label{dqqqd}
\end{equation}
\end{widetext}
The corresponding qubit representations of 
\begin{eqnarray}
    e^{-it\alpha_p n_p} &\rightarrow& e^{-i\frac{t\alpha_p}{2}} e^{i\frac{t\alpha_p}{2}\sigma_p^z}  \;, \label{eq127a} \\
    e^{-it\alpha_{pq} n_p n_q} &\rightarrow& e^{-i\frac{t\alpha_{pq}}{4} (1_p-\sigma_p^z)\otimes
    (1_q-\sigma_q^z)}  \;, \label{eq127b}
\end{eqnarray}
where $\alpha_p$ and $\alpha_{pq}$ are scalars, require two types of circuits (shown below respectively) that encode general 
$e^{-i\Theta \sigma_p^z}$ and $e^{-i\Theta \sigma_p^z\otimes\sigma_q^z}$ operators 
\begin{center}
\begin{quantikz}[transparent]
\lstick{$p$} & \gate{R_z(2\Theta)} & \qw
\end{quantikz} 
\end{center}
and 
\begin{center}
\begin{quantikz}[transparent]
\lstick{$p$} & \ctrl{1} & \qw                   & \ctrl{1} &  \qw \\ 
\lstick{$q$} & \targ{}  & \gate{R_z(2\Theta)} & \targ{}  &  \qw
\end{quantikz} 
\end{center}
which demonstrates the locality of the qubit encoding of the $e^{-it \widehat{\mathcal{P}}_E(G)}$ operator (or its approximate form defined by a truncated form of the  $\widehat{\mathcal{P}}_E(G)$ operator).

Further analysis of qubit encoding of (\ref{eq124}) requires explicit expansion  of 
$\widecheck{\mathcal{P}}(G)$ and $\widehat{\mathcal{P}}_M(G)$ in terms of Pauli strings $P_i$, i.e.,
\begin{eqnarray} 
\widecheck{\mathcal{P}}(G) &=& \sum_i \widecheck{g}_i P_i \;,\label{eq128} \\
\widehat{\mathcal{P}}_M(G) &=& \sum_j \widehat{h}_j P_j \;. \label{eq129}
\end{eqnarray}
where $\widecheck{g}_i$ and $\widehat{h}_j$ are scalars.
There is a simple way  how Trotter formula can be utilized to expand 
$e^{-it \lbrack \widecheck{\mathcal{P}}(G)+\widehat{\mathcal{P}}_M(G) \rbrack}$:
%\begin{itemize} 
% {$\boldsymbol{g-h}$ type:} 
%all $\widehat{h}_j P_j$ operators to the left of $\widecheck{g}_i P_i$ operators, i.e.,
\begin{equation}
 e^{-it \lbrack \widecheck{\mathcal{P}}(G)+\widehat{\mathcal{P}}_M(G) \rbrack}\simeq
 (X(r)Y(r))^r \;, \label{eq130}
\end{equation}
where 
\begin{eqnarray}
    X(r)&=&\prod_i e^{-i\frac{t\widecheck{g}_i}{r} P_i} \;. \label{eq131} \\
    Y(r)&=&\prod_j e^{-i\frac{t\widehat{h}_j}{r} P_j} \;, \label{eq131b} 
\end{eqnarray}
%\end{itemize}
According to  the definition of $\widehat{\mathcal{P}}_M(G)$ all Pauli strings in expansion (\ref{eq129}) are defined through the $Z$-gates, hence 
\begin{equation}
    Y(r)=e^{-i\frac{t}{r} \sum_j \widehat{h}_j P_j}\;.
    \label{eq132}
\end{equation}
Now, let's assume that $r$ is even and notice that  for sufficiently large $r$, the ordering  of the factors $X(r)$ and $Y(r)$  in (\ref{eq130}) can be arbitrary, i.e., it can be rewritten as 
\begin{eqnarray}
e^{-it \lbrack \widecheck{\mathcal{P}}(G)+\widehat{\mathcal{P}}_M(G) \rbrack} &\simeq& (X(r) Y(r) Y(r) X(r))^{\frac{r}{2}}
\label{133} \\
&=& (X(r) e^{-i\frac{2t}{r} \sum_j \widehat{h}_j P_j} X(r))^{\frac{r}{2}} \;.
\label{eq134} 
\end{eqnarray}
%the Trotter formula
%\begin{equation}
%    e^{-itG}\simeq (e^{-i\frac{t}{r} G})^r \;,
%    \label{eq126}
%\end{equation}
%can be simplified using the following steps:
%\begin{eqnarray}
%  (e^{-i\frac{t}{r} G})^r &=&  (e^{-i\frac{t}{r} \lbrack \widecheck{\mathcal{P}}(G)+\widehat{\mathcal{P}}_M(G)+\widehat{\mathcal{P}}_E(G) \rbrack})^r
%  \nonumber \\
%  &=&  (e^{-i\frac{t}{r} \lbrack \widecheck{\mathcal{P}}(G)+\widehat{\mathcal{P}}_M(G) \rbrack} e^{-i\frac{t}{r} \widehat{\mathcal{P}}_E(G)})^r \nonumber \\
%  &=& (e^{-i\frac{t}{r} \lbrack \widecheck{\mathcal{P}}(G)+\widehat{\mathcal{P}}_M(G) \rbrack})^r
%  e^{-it \widehat{\mathcal{P}}_E(G)} \;, \label{eq127} \\
%  &=& e^{-it \widehat{\mathcal{P}}_E(G)}
%  (e^{-i\frac{t}{r} \lbrack \widecheck{\mathcal{P}}(G)+\widehat{\mathcal{P}}_M(G) \rbrack})^r \;,
%  \label{eq127a}
%\end{eqnarray}
%which leads to an interesting form of the Trotter formula heavily using qubits corresponding to  the active spin-orbitals ($\widecheck{\mathcal{P}}(G)$ term), which are "weakly" coupled to the external qubits through the $\widehat{\mathcal{P}}_M(G)$ term. Additionally, the action of particle number operators can both stemming from  $\widehat{\mathcal{P}}_M(G)$ or $\widehat{\mathcal{P}}_E(G)$ in the exponents can be evaluated exactly. 

Combining (\ref{eq124},\ref{eq125b}) with (\ref{eq134}) leads to the further simplifications of the $\Omega(t)^{2^j}$:
\begin{equation}
 \Omega(t)^{2^j}\simeq 
 e^{-B}
 \lbrack 
 (X(r) 
 e^{-i\frac{2t}{r} \sum_j \widehat{h}_j P_j}
 X(r)
 )^{\frac{r}{2}}
 \rbrack^{2^j}
 e^{-it 2^j \widehat{\mathcal{P}}_E(G)} 
 e^B\;.
 \label{eq135}
\end{equation}
for sufficiently large values of $r$ parameter
(see Fig.\ref{fig5} instes (c) and (d)). 
\unskip
%
%
%\begin{figure}[H]
%\includegraphics[width=8.0 cm]{Fig5b.pdf}
%\caption{xxx yyy zzz}
%\label{fig5b}
%\end{figure}   
%\unskip
%%% k-locality
If the operator $B$ can be effectively calculated/approximated, the above formula offers several interesting properties:
\begin{itemize}
    \item all terms depending on the $\widehat{\mathcal{P}}_M(G)$ and $\widehat{\mathcal{P}}_E(G)$ can be evaluated exactly,
    \item the $\widecheck{\mathcal{P}}(G)$ is $2(n-k)$-local (in the sense of Ref.\cite{lloyd1996universal}) i.e., is defined interactions involving at most  $2(n-k)$ qubits (assuming qubit ordering defined in scheme \ref{eq7}),
    \item for $(n-k)\ll k$, if the  operator $B$ can be approximated by single and double excitations (approach consistent with the first order of perturbation theory), then the number of terms that need to be included in the $B$ operator is proportional to  $ (n-k)\times k^3$ (smaller number of terms compared to the original Hamiltonian $H$, which is proportional to $n^4$). Additionally, there are only two instances when the $e^{-B}$/$e^B$ transformations have to be performed to encode $\Omega(t)^{2j}$ (see, Eq.(\ref{eq124})).
    \item expansion (\ref{eq135}) is reminiscent of the QAOA (Quantum Approximate Optimization Algorithm)  Ansatz \cite{farhi2014quantum,farhi2017quantum,zhu2020adaptive,kremenetski2021quantum,herrman2022multi}
    consisting of alternating  sequence of  cost ($\widehat{\mathcal{P}}_M(G)$-dependent part) and mixer ($\widecheck{\mathcal{P}}(G)$-dependent part) layers.
\end{itemize}
%For the approximate models where $G$ operator is for example approximated by retaining only  one- and two-body interactions the RRST representation can provide significant savings in terms of the circuit depth and circuit complexity and additionally provide access to multiple-states for several systems characterized by various numbers of electron that can be jointly approximated by the same active spin-orbitals. 
Also, in analogy to the Fock-space coupled-cluster formulations (for example, the valence-universal (VU-CC) theories), the
$B$ and $G$ operators along with the corresponding time evolution operator $\Omega(t)$ can act on the states $|\Theta(n_e)\rangle$ corresponding to various number of  electron  ($n_e$), i.e.,
\begin{equation}
    \Omega(t)^{2^j} |\Theta(n_e)\rangle \;,
    \label{eq1129}
\end{equation}
providing access to energies of neutral, ionized, doubly-ionized, etc., electronic states having non-zero overlap with $|\Theta(n_e)\rangle$. A similar algorithm can be used to evaluate phase using representation (\ref{eq29}) based on the auxiliary unitary transformation.

%%--%%--%%
%\begin{figure}[H]
%\includegraphics[width=8.5 cm]{RRUST_Fig1.pdf}
%\caption{A comparison of schematic representations of quantum circuit needed to form powers of the evolution operators $\Omega(t)=e^{-itH}$. (a) General structure of the QPE algorithm. (b) Powers of the evolution operator using standard representation of Hamiltonian $H$\label{fig2}}
%\end{figure}   
%\unskip
%%%%%%%%%%%%

% OKKK

\section{Approximations}

The potential of reducing hardware requirements discussed in the previous Section  is inextricably connected to our ability to approximate solutions of non-terminating expansions (\ref{eq13}) using their finite-rank commutator expansion (\ref{eq14}). We envision this step to be entirely performed using conventional computers. This step is justified because the process of solving  equations for finite-rank commutator expansions is associated with polynomial scaling. 

In analogy to the existing quantum algorithms, including a broad class of VQE and QPE formulations, classes of approximations are indispensable to translate the problem into the form of circuits for quantum computers or their classical emulators. 
For example, the broad class of Trotter-based approximations is needed for realizing unitary CC-driven variants of the VQE formalism and evolution operators in the QPE formalism.
It does not come as a surprise that for the SW-RRST approximation, the situation is no different.  
The main factors that need to be taken into account when defining approximate SW-RRST formulations are as follows:
\begin{enumerate}
    \item {\bf The excitation-rank of the many-body form of the  transformed $G$ Hamiltonian.} Our experience with Hilbert-space-type downfolding indicates that one- and two-body effective interactions can provide satisfactory results when active space is adequately defined.
    \cite{bauman2022coupled2c}. 
%    However, one should be aware that the Hilbert-space downfolding uses mixed particle-hole and  Fermi-vaccum  formalisms, that when combined can effectively compress large class of correlation effects in the effective Hamiltonians.
    \item {\bf The rank of the many-body effects in the $B$ operator.} The elementary perturbative analysis for the case discussed in Corollary 1 indicates that the 0-th order of $B$ is equal to zero, the 1-st order contributed to one- and two-body terms, while the 2-nd order introduces higher-rank effects
    (see the analysis in Appendix A). It suggests that simple models based on the inclusion of  one- and two-body  effects in the $B$ operator are justified. 
    \item {\bf The working equations for $B$-amplitudes.} Due to their non-terminating nature, the algebraic form of  equations (\ref{eq22}) or (\ref{eq26}) has to be approximated due to their non-terminating nature. A numerically feasible way of introducing sufficiency conditions for $B$-operator  amplitudes is to use finite commutator-rank expansion in  Eqs. (\ref{eq22}) or (\ref{eq26}). The class of approximations termed SW-RRST($l$) consist in retaining commutators up to the $l$-th rank. For formulation (\ref{eq22}) we have
\begin{equation}
  \widehat{\mathcal{P}}_{eod}(H +  \sum_{i=1}^{l} (-1)^i \frac{1}{i!}[
   \ldots [H,B],\ldots ],B]_i
   ) =0   \;.
   \label{eq100}
\end{equation}
    whereas for (\ref{eq26}) one gets
\begin{equation}
  \widehat{\mathcal{P}}_{od}(H +  \sum_{i=1}^{l} (-1)^i \frac{1}{i!}[
   \ldots [H,B],\ldots ],B]_i
   ) =0   \;.
   \label{eq101}
\end{equation}
While the linear SW-RRST(1) is the simplest approximation consistent with the low-order perturbative analysis, one may have to  deal with the singular or nearly singular forms of the equation.  Similar problems have been observed in the early studies of the  CC theory. \cite{jankowski1980applicability} 
Several techniques akin to almost-linear CC approximations are discussed  in Refs. \cite{jankowski1996approximate,jankowski1998approximate,li2000approximate}, including 
the auxiliary transformation discussed earlier,
have the potential to  offset singular behavior. Another factor that was shown to play a key role in the removal of singularities of linear CC equation is associated with the inclusion of non-linear terms, which is the main motivation for the development of higher-rank SW-RRST(l) approximations ($l > 1$).
\end{enumerate}
Another important aspect of the proposed approximation is the choice of the orbital basis that should be driven by the targeted quantum system. 

% OKKK

%%%%%%%%%%%%%%%%%%%%%%%%%%%%%%%%%%%%%%%%%%
\section{Conclusions}
In summary, we have discussed the extension of the traditional 
downfolding methods to the Fock-space formulation using the Schrieffer--Wolff-type transformations. We have presented the basic properties of this formalism that leads to a simplified form of the similarity-transformed Hamiltonian $G$ that, in the context of the quantum circuit complexity, is dominated by its $2(n-k)$-local internal part. The remaining external part of the $G$ operator, depending on the specific form of the approach, either given by conditions (\ref{eq22}) or (\ref{eq26}), is defined by simple operators (for example, particle number operators) that can be determined exactly. 
An additional advantage of the SW-RRST approach is that the external components of the $G$ operator can be factored out and do not need to be handled by the Trotter expansion. 
While we have discussed the general features of the SW-RRST formalism and its approximation, in the following papers, we will analyze the numerical solutions to the SW-RRST($l$) equations and the effect of non-linear terms on the amplitudes defining the anti-Hermitian $B$ operator. An essential aspect of the numerical analysis is the determination of the feasibility of auxiliary similarity transformation that breaks symmetries of the targeted quantum system. 

%%%%%%%%%%%%%%%%%%%%%%%%%%%%%%%%%%%%%%%%%%

%%%%%%%%%%%%%%%%%%%%%%%%%%%%%%%%%%%%%%%%%%
\vspace{6pt} 

%%%%%%%%%%%%%%%%%%%%%%%%%%%%%%%%%%%%%%%%%%
%% optional
%\supplementary{The following supporting information can be downloaded at:  \linksupplementary{s1}, Figure S1: title; Table S1: title; Video S1: title.}


%%%%%%%%%%%%%%%%%%%%%%%%%%%%%%%%%%%%%%%%%%

\section{Acknowledgement}
This  work  was  supported  by  the ``Embedding Quantum Computing into Many-body Frameworks for Strongly Correlated  Molecular and Materials Systems'' project, which is funded by the U.S. Department of Energy(DOE), Office of Science, Office of Basic Energy Sciences, the Division of Chemical Sciences, Geosciences, and Biosciences. 
The work was performed at the Pacific Northwest National Laboratory (PNNL). 
PNNL is operated for the U.S. Department of Energy by the Battelle Memorial Institute under Contract DE-AC06-76RLO-1830.

%\appendixstart
\appendix
%%\section[\appendixname~\thesection]{}
\section[\appendixname~\thesection]{Perturbative estimates of the $B$ operator}

Our perturbative analysis of the $B$ operator will utilize the following partitioning of the Hamiltonian (\ref{eq2})
into 0-th order part $H_0$ (assumed to be diagonal) and perturbation $W$
\begin{equation}
    H(\lambda) =H_0+
    \lambda W \;
    \label{app1}
\end{equation}
where $H_0$ and $W$ are generally defined as
\begin{equation}
    H_0=\sum_{p,q} (h^p_q-v^p_q) a_p^{\dagger}a_q 
    =\sum_{p} \epsilon_p a_p^{\dagger} a_p
    \label{app2}
\end{equation}
and 
\begin{eqnarray}
   W&=&V_1+V_2 \;, \label{app3} \\
   V_1 &=& \sum_{p,q} v^p_q a_p^{\dagger}a_q \;, \label{app4} \\
   V_2 &=& \frac{1}{4} \sum_{p,q,r,s}
v^{pq}_{rs} a_p^{\dagger} a_q^{\dagger} a_s a_r \;. \label{app5} 
\end{eqnarray}
At this point, we assume the diagonal form of the 
$H_0$ operator without specifying the form of the $V_1$ operator, which can generally be a spatial/spin symmetry-breaking operator.
In particular, we are not assuming that $\epsilon_p$'s are Hartree-Fock spin-orbital energies, which would be challenging in situations when Hartree-Fock  external orbitals are degenerate. These problems can  be addressed by the proper definition of $H_0$ (or by the properly designed  gauge transformation mentioned earlier). 

For simplicity, we assume that the perturbative expansion for the $B$ operator is convergent  and  takes the form 
\begin{equation}
B=\sum_{i=0}^{\infty} \lambda^i B^{(i)} \;,
\label{app6}
\end{equation}
where $i$ refers to the order of perturbative expansion. 
Introducing (\ref{app6}) into sufficiency conditions (\ref{eq22}) (similar analysis is valid for the variant described in Corollary 1 or for the Hamiltonian $\bar{H}_C$ (\ref{eq30}) ) 
\begin{equation}
\widehat{\mathcal{P}}_{eod}
(e^{\sum_{i=0}\lambda^i B^{(i)}}
(H_0+\lambda W)
e^{-\sum_{j=0}\lambda^j B^{(j)}}) 
=0 \;,
\label{app7}
\end{equation}
we get, using BCH expansion, equations for perturbative components of the $B$ operator. For example, 
\begin{itemize}
\item {\bf 0-th order:}
    \begin{equation}
    \widehat{\mathcal{P}}_{eod}([H_0,B^{(0)}]) =0 \;,
     \label{app8}
    \end{equation}
    where we utilized the fact that $H_0$ is a diagonal operator 
    ($\widehat{\mathcal{P}}_{eod}(H_0)=0$), which leads to $B^{(0)}=0$.
\item {\bf 1-st order:}
\begin{equation}
      \widehat{\mathcal{P}}_{eod}(W-[H_0,B^{(1)}]) =0 \;,
     \label{app9}
\end{equation}
which yields  one- and two-body components only. 
\item {\bf 2-nd order:}
\begin{equation}
      \widehat{\mathcal{P}}_{eod}(-[H_0,B^{(2)}]
      +\frac{1}{2}[[H_0,B^{(1)}],B^{(1)}]
      -[W,B^{(1)}])
      =0 \;,
     \label{app9b}
\end{equation}
which generates lowest-order three-body interactions
(stemming from the $[W,B^{(1)}]$ term). 
\end{itemize}
As a consequence of the non-linear character of the expansion (\ref{app7}), higher orders of perturbation theory generate higher-rank many-body components. It is interesting to notice that the rank of excitation  vs. its perturbation order is essentially the same as in the standard SR-CC theory.

%\bibliography{ref3.bib}

\begin{thebibliography}{107}%
\makeatletter
\providecommand \@ifxundefined [1]{%
 \@ifx{#1\undefined}
}%
\providecommand \@ifnum [1]{%
 \ifnum #1\expandafter \@firstoftwo
 \else \expandafter \@secondoftwo
 \fi
}%
\providecommand \@ifx [1]{%
 \ifx #1\expandafter \@firstoftwo
 \else \expandafter \@secondoftwo
 \fi
}%
\providecommand \natexlab [1]{#1}%
\providecommand \enquote  [1]{``#1''}%
\providecommand \bibnamefont  [1]{#1}%
\providecommand \bibfnamefont [1]{#1}%
\providecommand \citenamefont [1]{#1}%
\providecommand \href@noop [0]{\@secondoftwo}%
\providecommand \href [0]{\begingroup \@sanitize@url \@href}%
\providecommand \@href[1]{\@@startlink{#1}\@@href}%
\providecommand \@@href[1]{\endgroup#1\@@endlink}%
\providecommand \@sanitize@url [0]{\catcode `\\12\catcode `\$12\catcode
  `\&12\catcode `\#12\catcode `\^12\catcode `\_12\catcode `\%12\relax}%
\providecommand \@@startlink[1]{}%
\providecommand \@@endlink[0]{}%
\providecommand \url  [0]{\begingroup\@sanitize@url \@url }%
\providecommand \@url [1]{\endgroup\@href {#1}{\urlprefix }}%
\providecommand \urlprefix  [0]{URL }%
\providecommand \Eprint [0]{\href }%
\providecommand \doibase [0]{http://dx.doi.org/}%
\providecommand \selectlanguage [0]{\@gobble}%
\providecommand \bibinfo  [0]{\@secondoftwo}%
\providecommand \bibfield  [0]{\@secondoftwo}%
\providecommand \translation [1]{[#1]}%
\providecommand \BibitemOpen [0]{}%
\providecommand \bibitemStop [0]{}%
\providecommand \bibitemNoStop [0]{.\EOS\space}%
\providecommand \EOS [0]{\spacefactor3000\relax}%
\providecommand \BibitemShut  [1]{\csname bibitem#1\endcsname}%
\let\auto@bib@innerbib\@empty
%</preamble>
\bibitem [{\citenamefont {Coester}(1958)}]{coester58_421}%
  \BibitemOpen
  \bibfield  {author} {\bibinfo {author} {\bibfnamefont {F.}~\bibnamefont
  {Coester}},\ }\href {\doibase http://dx.doi.org/10.1016/0029-5582(58)90280-3}
  {\bibfield  {journal} {\bibinfo  {journal} {Nucl. Phys.}\ }\textbf {\bibinfo
  {volume} {7}},\ \bibinfo {pages} {421} (\bibinfo {year} {1958})}\BibitemShut
  {NoStop}%
\bibitem [{\citenamefont {Coester}\ and\ \citenamefont
  {Kummel}(1960)}]{coester60_477}%
  \BibitemOpen
  \bibfield  {author} {\bibinfo {author} {\bibfnamefont {F.}~\bibnamefont
  {Coester}}\ and\ \bibinfo {author} {\bibfnamefont {H.}~\bibnamefont
  {Kummel}},\ }\href {\doibase http://dx.doi.org/10.1016/0029-5582(60)90140-1}
  {\bibfield  {journal} {\bibinfo  {journal} {Nucl. Phys.}\ }\textbf {\bibinfo
  {volume} {17}},\ \bibinfo {pages} {477} (\bibinfo {year} {1960})}\BibitemShut
  {NoStop}%
\bibitem [{\citenamefont {{\v C}{\'\i}{\v z}ek}(1966)}]{cizek66_4256}%
  \BibitemOpen
  \bibfield  {author} {\bibinfo {author} {\bibfnamefont {J.}~\bibnamefont {{\v
  C}{\'\i}{\v z}ek}},\ }\href {\doibase http://dx.doi.org/10.1063/1.1727484}
  {\bibfield  {journal} {\bibinfo  {journal} {J. Chem. Phys.}\ }\textbf
  {\bibinfo {volume} {45}},\ \bibinfo {pages} {4256} (\bibinfo {year}
  {1966})}\BibitemShut {NoStop}%
\bibitem [{\citenamefont {Paldus}, \citenamefont {\ifmmode \check{C}\else
  \v{C}\fi{}\'{\i}\ifmmode~\check{z}\else \v{z}\fi{}ek},\ and\ \citenamefont
  {Shavitt}(1972)}]{paldus72_50}%
  \BibitemOpen
  \bibfield  {author} {\bibinfo {author} {\bibfnamefont {J.}~\bibnamefont
  {Paldus}}, \bibinfo {author} {\bibfnamefont {J.}~\bibnamefont {\ifmmode
  \check{C}\else \v{C}\fi{}\'{\i}\ifmmode~\check{z}\else \v{z}\fi{}ek}}, \ and\
  \bibinfo {author} {\bibfnamefont {I.}~\bibnamefont {Shavitt}},\ }\href
  {\doibase 10.1103/PhysRevA.5.50} {\bibfield  {journal} {\bibinfo  {journal}
  {Phys. Rev. A}\ }\textbf {\bibinfo {volume} {5}},\ \bibinfo {pages} {50}
  (\bibinfo {year} {1972})}\BibitemShut {NoStop}%
\bibitem [{\citenamefont {Purvis}\ and\ \citenamefont
  {Bartlett}(1982)}]{purvis82_1910}%
  \BibitemOpen
  \bibfield  {author} {\bibinfo {author} {\bibfnamefont {G.~D.}\ \bibnamefont
  {Purvis}}\ and\ \bibinfo {author} {\bibfnamefont {R.~J.}\ \bibnamefont
  {Bartlett}},\ }\href {\doibase http://dx.doi.org/10.1063/1.443164} {\bibfield
   {journal} {\bibinfo  {journal} {J. Chem. Phys.}\ }\textbf {\bibinfo {volume}
  {76}},\ \bibinfo {pages} {1910} (\bibinfo {year} {1982})}\BibitemShut
  {NoStop}%
\bibitem [{\citenamefont {Arponen}(1983)}]{arponen83_311}%
  \BibitemOpen
  \bibfield  {author} {\bibinfo {author} {\bibfnamefont {J.}~\bibnamefont
  {Arponen}},\ }\href {\doibase http://dx.doi.org/10.1016/0003-4916(83)90284-1}
  {\bibfield  {journal} {\bibinfo  {journal} {Ann. Phys.}\ }\textbf {\bibinfo
  {volume} {151}},\ \bibinfo {pages} {311} (\bibinfo {year}
  {1983})}\BibitemShut {NoStop}%
\bibitem [{\citenamefont {Bishop}(1991)}]{bishop1991overview}%
  \BibitemOpen
  \bibfield  {author} {\bibinfo {author} {\bibfnamefont {R.}~\bibnamefont
  {Bishop}},\ }\href@noop {} {\bibfield  {journal} {\bibinfo  {journal}
  {Theoretica chimica acta}\ }\textbf {\bibinfo {volume} {80}},\ \bibinfo
  {pages} {95} (\bibinfo {year} {1991})}\BibitemShut {NoStop}%
\bibitem [{\citenamefont {Koch}\ and\ \citenamefont
  {J{\o}rgensen}(1990)}]{jorgensen90_3333}%
  \BibitemOpen
  \bibfield  {author} {\bibinfo {author} {\bibfnamefont {H.}~\bibnamefont
  {Koch}}\ and\ \bibinfo {author} {\bibfnamefont {P.}~\bibnamefont
  {J{\o}rgensen}},\ }\href {\doibase 10.1063/1.458814} {\bibfield  {journal}
  {\bibinfo  {journal} {J. Chem. Phys.}\ }\textbf {\bibinfo {volume} {93}},\
  \bibinfo {pages} {3333} (\bibinfo {year} {1990})}\BibitemShut {NoStop}%
\bibitem [{\citenamefont {Paldus}\ and\ \citenamefont {Li}(1999)}]{paldus07}%
  \BibitemOpen
  \bibfield  {author} {\bibinfo {author} {\bibfnamefont {J.}~\bibnamefont
  {Paldus}}\ and\ \bibinfo {author} {\bibfnamefont {X.}~\bibnamefont {Li}},\
  }\href {\doibase 10.1002/9780470141694.ch1} {\bibfield  {journal} {\bibinfo
  {journal} {Adv. Chem. Phys.}\ }\textbf {\bibinfo {volume} {110}},\ \bibinfo
  {pages} {1} (\bibinfo {year} {1999})}\BibitemShut {NoStop}%
\bibitem [{\citenamefont {Crawford}\ and\ \citenamefont
  {Schaefer}(2000)}]{crawford2000introduction}%
  \BibitemOpen
  \bibfield  {author} {\bibinfo {author} {\bibfnamefont {T.~D.}\ \bibnamefont
  {Crawford}}\ and\ \bibinfo {author} {\bibfnamefont {H.~F.}\ \bibnamefont
  {Schaefer}},\ }\href@noop {} {\bibfield  {journal} {\bibinfo  {journal} {Rev.
  Comput. Chem.}\ }\textbf {\bibinfo {volume} {14}},\ \bibinfo {pages} {33}
  (\bibinfo {year} {2000})}\BibitemShut {NoStop}%
\bibitem [{\citenamefont {Bartlett}\ and\ \citenamefont
  {Musia\l}(2007)}]{bartlett_rmp}%
  \BibitemOpen
  \bibfield  {author} {\bibinfo {author} {\bibfnamefont {R.~J.}\ \bibnamefont
  {Bartlett}}\ and\ \bibinfo {author} {\bibfnamefont {M.}~\bibnamefont
  {Musia\l}},\ }\href {\doibase 10.1103/RevModPhys.79.291} {\bibfield
  {journal} {\bibinfo  {journal} {Rev. Mod. Phys.}\ }\textbf {\bibinfo {volume}
  {79}},\ \bibinfo {pages} {291} (\bibinfo {year} {2007})}\BibitemShut
  {NoStop}%
\bibitem [{\citenamefont {Arponen}, \citenamefont {Bishop},\ and\ \citenamefont
  {Pajanne}(1987{\natexlab{a}})}]{arponen1987extended1}%
  \BibitemOpen
  \bibfield  {author} {\bibinfo {author} {\bibfnamefont {J.}~\bibnamefont
  {Arponen}}, \bibinfo {author} {\bibfnamefont {R.}~\bibnamefont {Bishop}}, \
  and\ \bibinfo {author} {\bibfnamefont {E.}~\bibnamefont {Pajanne}},\
  }\href@noop {} {\bibfield  {journal} {\bibinfo  {journal} {Physical Review
  A}\ }\textbf {\bibinfo {volume} {36}},\ \bibinfo {pages} {2519} (\bibinfo
  {year} {1987}{\natexlab{a}})}\BibitemShut {NoStop}%
\bibitem [{\citenamefont {Arponen}, \citenamefont {Bishop},\ and\ \citenamefont
  {Pajanne}(1987{\natexlab{b}})}]{arponen1987extended}%
  \BibitemOpen
  \bibfield  {author} {\bibinfo {author} {\bibfnamefont {J.}~\bibnamefont
  {Arponen}}, \bibinfo {author} {\bibfnamefont {R.}~\bibnamefont {Bishop}}, \
  and\ \bibinfo {author} {\bibfnamefont {E.}~\bibnamefont {Pajanne}},\
  }\href@noop {} {\bibfield  {journal} {\bibinfo  {journal} {Physical Review
  A}\ }\textbf {\bibinfo {volume} {36}},\ \bibinfo {pages} {2539} (\bibinfo
  {year} {1987}{\natexlab{b}})}\BibitemShut {NoStop}%
\bibitem [{\citenamefont {Arponen}\ and\ \citenamefont
  {Bishop}(1991{\natexlab{a}})}]{arponen1991independent}%
  \BibitemOpen
  \bibfield  {author} {\bibinfo {author} {\bibfnamefont {J.}~\bibnamefont
  {Arponen}}\ and\ \bibinfo {author} {\bibfnamefont {R.}~\bibnamefont
  {Bishop}},\ }\href@noop {} {\bibfield  {journal} {\bibinfo  {journal} {Annals
  of Physics}\ }\textbf {\bibinfo {volume} {207}},\ \bibinfo {pages} {171}
  (\bibinfo {year} {1991}{\natexlab{a}})}\BibitemShut {NoStop}%
\bibitem [{\citenamefont {Arponen}\ and\ \citenamefont
  {Bishop}(1993)}]{arponen1993independent}%
  \BibitemOpen
  \bibfield  {author} {\bibinfo {author} {\bibfnamefont {J.}~\bibnamefont
  {Arponen}}\ and\ \bibinfo {author} {\bibfnamefont {R.}~\bibnamefont
  {Bishop}},\ }\href@noop {} {\bibfield  {journal} {\bibinfo  {journal} {Annals
  of Physics}\ }\textbf {\bibinfo {volume} {227}},\ \bibinfo {pages} {275}
  (\bibinfo {year} {1993})}\BibitemShut {NoStop}%
\bibitem [{\citenamefont {Robinson}, \citenamefont {Bishop},\ and\
  \citenamefont {Arponen}(1989)}]{robinson1989extended}%
  \BibitemOpen
  \bibfield  {author} {\bibinfo {author} {\bibfnamefont {N.}~\bibnamefont
  {Robinson}}, \bibinfo {author} {\bibfnamefont {R.}~\bibnamefont {Bishop}}, \
  and\ \bibinfo {author} {\bibfnamefont {J.}~\bibnamefont {Arponen}},\
  }\href@noop {} {\bibfield  {journal} {\bibinfo  {journal} {Physical Review
  A}\ }\textbf {\bibinfo {volume} {40}},\ \bibinfo {pages} {4256} (\bibinfo
  {year} {1989})}\BibitemShut {NoStop}%
\bibitem [{\citenamefont {Arponen}\ and\ \citenamefont
  {Bishop}(1991{\natexlab{b}})}]{arponen1991holomorphic}%
  \BibitemOpen
  \bibfield  {author} {\bibinfo {author} {\bibfnamefont {J.~S.}\ \bibnamefont
  {Arponen}}\ and\ \bibinfo {author} {\bibfnamefont {R.~F.}\ \bibnamefont
  {Bishop}},\ }\href@noop {} {\bibfield  {journal} {\bibinfo  {journal}
  {Theoretica chimica acta}\ }\textbf {\bibinfo {volume} {80}},\ \bibinfo
  {pages} {289} (\bibinfo {year} {1991}{\natexlab{b}})}\BibitemShut {NoStop}%
\bibitem [{\citenamefont {Emrich}\ and\ \citenamefont
  {Zabolitzky}(1984)}]{emrich1984electron}%
  \BibitemOpen
  \bibfield  {author} {\bibinfo {author} {\bibfnamefont {K.}~\bibnamefont
  {Emrich}}\ and\ \bibinfo {author} {\bibfnamefont {J.}~\bibnamefont
  {Zabolitzky}},\ }\href@noop {} {\bibfield  {journal} {\bibinfo  {journal}
  {Physical Review B}\ }\textbf {\bibinfo {volume} {30}},\ \bibinfo {pages}
  {2049} (\bibinfo {year} {1984})}\BibitemShut {NoStop}%
\bibitem [{\citenamefont {Funke}, \citenamefont {Kaulfuss},\ and\ \citenamefont
  {K{\"u}mmel}(1987)}]{funke1987approaching}%
  \BibitemOpen
  \bibfield  {author} {\bibinfo {author} {\bibfnamefont {M.}~\bibnamefont
  {Funke}}, \bibinfo {author} {\bibfnamefont {U.}~\bibnamefont {Kaulfuss}}, \
  and\ \bibinfo {author} {\bibfnamefont {H.}~\bibnamefont {K{\"u}mmel}},\
  }\href@noop {} {\bibfield  {journal} {\bibinfo  {journal} {Physical Review
  D}\ }\textbf {\bibinfo {volume} {35}},\ \bibinfo {pages} {621} (\bibinfo
  {year} {1987})}\BibitemShut {NoStop}%
\bibitem [{\citenamefont {K{\"u}mmel}(2001)}]{kummel2001post}%
  \BibitemOpen
  \bibfield  {author} {\bibinfo {author} {\bibfnamefont {H.~G.}\ \bibnamefont
  {K{\"u}mmel}},\ }\href@noop {} {\bibfield  {journal} {\bibinfo  {journal}
  {Physical Review B}\ }\textbf {\bibinfo {volume} {64}},\ \bibinfo {pages}
  {014301} (\bibinfo {year} {2001})}\BibitemShut {NoStop}%
\bibitem [{\citenamefont {Hasberg}\ and\ \citenamefont
  {K{\"u}mmel}(1986)}]{hasberg1986coupled}%
  \BibitemOpen
  \bibfield  {author} {\bibinfo {author} {\bibfnamefont {G.}~\bibnamefont
  {Hasberg}}\ and\ \bibinfo {author} {\bibfnamefont {H.}~\bibnamefont
  {K{\"u}mmel}},\ }\href@noop {} {\bibfield  {journal} {\bibinfo  {journal}
  {Physical Review C}\ }\textbf {\bibinfo {volume} {33}},\ \bibinfo {pages}
  {1367} (\bibinfo {year} {1986})}\BibitemShut {NoStop}%
\bibitem [{\citenamefont {Bishop}, \citenamefont {Ligterink},\ and\
  \citenamefont {Walet}(2006)}]{bishop2006towards}%
  \BibitemOpen
  \bibfield  {author} {\bibinfo {author} {\bibfnamefont {R.~F.}\ \bibnamefont
  {Bishop}}, \bibinfo {author} {\bibfnamefont {N.}~\bibnamefont {Ligterink}}, \
  and\ \bibinfo {author} {\bibfnamefont {N.~R.}\ \bibnamefont {Walet}},\
  }\href@noop {} {\bibfield  {journal} {\bibinfo  {journal} {International
  journal of modern physics B}\ }\textbf {\bibinfo {volume} {20}},\ \bibinfo
  {pages} {4992} (\bibinfo {year} {2006})}\BibitemShut {NoStop}%
\bibitem [{\citenamefont {Ligterink}, \citenamefont {Walet},\ and\
  \citenamefont {Bishop}(1998)}]{ligterink1998coupled}%
  \BibitemOpen
  \bibfield  {author} {\bibinfo {author} {\bibfnamefont {N.}~\bibnamefont
  {Ligterink}}, \bibinfo {author} {\bibfnamefont {N.}~\bibnamefont {Walet}}, \
  and\ \bibinfo {author} {\bibfnamefont {R.}~\bibnamefont {Bishop}},\
  }\href@noop {} {\bibfield  {journal} {\bibinfo  {journal} {Annals of
  Physics}\ }\textbf {\bibinfo {volume} {267}},\ \bibinfo {pages} {97}
  (\bibinfo {year} {1998})}\BibitemShut {NoStop}%
\bibitem [{\citenamefont {Arponen}\ \emph {et~al.}(1988)\citenamefont
  {Arponen}, \citenamefont {Bishop}, \citenamefont {Pajanne},\ and\
  \citenamefont {Robinson}}]{arponen1988towards}%
  \BibitemOpen
  \bibfield  {author} {\bibinfo {author} {\bibfnamefont {J.}~\bibnamefont
  {Arponen}}, \bibinfo {author} {\bibfnamefont {R.}~\bibnamefont {Bishop}},
  \bibinfo {author} {\bibfnamefont {E.}~\bibnamefont {Pajanne}}, \ and\
  \bibinfo {author} {\bibfnamefont {N.}~\bibnamefont {Robinson}},\ }in\
  \href@noop {} {\emph {\bibinfo {booktitle} {Condensed matter theories}}}\
  (\bibinfo  {publisher} {Springer},\ \bibinfo {year} {1988})\ pp.\ \bibinfo
  {pages} {51--66}\BibitemShut {NoStop}%
\bibitem [{\citenamefont {Bishop}\ \emph {et~al.}(1989)\citenamefont {Bishop},
  \citenamefont {Robinson}, \citenamefont {Arponen},\ and\ \citenamefont
  {Pajanne}}]{bishop1989quantum}%
  \BibitemOpen
  \bibfield  {author} {\bibinfo {author} {\bibfnamefont {R.}~\bibnamefont
  {Bishop}}, \bibinfo {author} {\bibfnamefont {N.}~\bibnamefont {Robinson}},
  \bibinfo {author} {\bibfnamefont {J.}~\bibnamefont {Arponen}}, \ and\
  \bibinfo {author} {\bibfnamefont {E.}~\bibnamefont {Pajanne}},\ }in\
  \href@noop {} {\emph {\bibinfo {booktitle} {Aspects of Many-Body Effects in
  Molecules and Extended Systems}}}\ (\bibinfo  {publisher} {Springer},\
  \bibinfo {year} {1989})\ pp.\ \bibinfo {pages} {241--260}\BibitemShut
  {NoStop}%
\bibitem [{\citenamefont {Dean}\ and\ \citenamefont
  {Hjorth-Jensen}(2004)}]{PhysRevC.69.054320}%
  \BibitemOpen
  \bibfield  {author} {\bibinfo {author} {\bibfnamefont {D.~J.}\ \bibnamefont
  {Dean}}\ and\ \bibinfo {author} {\bibfnamefont {M.}~\bibnamefont
  {Hjorth-Jensen}},\ }\href {\doibase 10.1103/PhysRevC.69.054320} {\bibfield
  {journal} {\bibinfo  {journal} {Phys. Rev. C}\ }\textbf {\bibinfo {volume}
  {69}},\ \bibinfo {pages} {054320} (\bibinfo {year} {2004})}\BibitemShut
  {NoStop}%
\bibitem [{\citenamefont {Kowalski}\ \emph {et~al.}(2004)\citenamefont
  {Kowalski}, \citenamefont {Dean}, \citenamefont {Hjorth-Jensen},
  \citenamefont {Papenbrock},\ and\ \citenamefont
  {Piecuch}}]{PhysRevLett.92.132501}%
  \BibitemOpen
  \bibfield  {author} {\bibinfo {author} {\bibfnamefont {K.}~\bibnamefont
  {Kowalski}}, \bibinfo {author} {\bibfnamefont {D.~J.}\ \bibnamefont {Dean}},
  \bibinfo {author} {\bibfnamefont {M.}~\bibnamefont {Hjorth-Jensen}}, \bibinfo
  {author} {\bibfnamefont {T.}~\bibnamefont {Papenbrock}}, \ and\ \bibinfo
  {author} {\bibfnamefont {P.}~\bibnamefont {Piecuch}},\ }\href {\doibase
  10.1103/PhysRevLett.92.132501} {\bibfield  {journal} {\bibinfo  {journal}
  {Phys. Rev. Lett.}\ }\textbf {\bibinfo {volume} {92}},\ \bibinfo {pages}
  {132501} (\bibinfo {year} {2004})}\BibitemShut {NoStop}%
\bibitem [{\citenamefont {Hagen}\ \emph {et~al.}(2008)\citenamefont {Hagen},
  \citenamefont {Papenbrock}, \citenamefont {Dean},\ and\ \citenamefont
  {Hjorth-Jensen}}]{PhysRevLett.101.092502}%
  \BibitemOpen
  \bibfield  {author} {\bibinfo {author} {\bibfnamefont {G.}~\bibnamefont
  {Hagen}}, \bibinfo {author} {\bibfnamefont {T.}~\bibnamefont {Papenbrock}},
  \bibinfo {author} {\bibfnamefont {D.~J.}\ \bibnamefont {Dean}}, \ and\
  \bibinfo {author} {\bibfnamefont {M.}~\bibnamefont {Hjorth-Jensen}},\ }\href
  {\doibase 10.1103/PhysRevLett.101.092502} {\bibfield  {journal} {\bibinfo
  {journal} {Phys. Rev. Lett.}\ }\textbf {\bibinfo {volume} {101}},\ \bibinfo
  {pages} {092502} (\bibinfo {year} {2008})}\BibitemShut {NoStop}%
\bibitem [{\citenamefont {Scheiner}\ \emph {et~al.}(1987)\citenamefont
  {Scheiner}, \citenamefont {Scuseria}, \citenamefont {Rice}, \citenamefont
  {Lee},\ and\ \citenamefont {Schaefer~III}}]{scheiner1987analytic}%
  \BibitemOpen
  \bibfield  {author} {\bibinfo {author} {\bibfnamefont {A.~C.}\ \bibnamefont
  {Scheiner}}, \bibinfo {author} {\bibfnamefont {G.~E.}\ \bibnamefont
  {Scuseria}}, \bibinfo {author} {\bibfnamefont {J.~E.}\ \bibnamefont {Rice}},
  \bibinfo {author} {\bibfnamefont {T.~J.}\ \bibnamefont {Lee}}, \ and\
  \bibinfo {author} {\bibfnamefont {H.~F.}\ \bibnamefont {Schaefer~III}},\
  }\href@noop {} {\bibfield  {journal} {\bibinfo  {journal} {J. Chem. Phys.}\
  }\textbf {\bibinfo {volume} {87}},\ \bibinfo {pages} {5361} (\bibinfo {year}
  {1987})}\BibitemShut {NoStop}%
\bibitem [{\citenamefont {Sinnokrot}, \citenamefont {Valeev},\ and\
  \citenamefont {Sherrill}(2002)}]{sinnokrot2002estimates}%
  \BibitemOpen
  \bibfield  {author} {\bibinfo {author} {\bibfnamefont {M.~O.}\ \bibnamefont
  {Sinnokrot}}, \bibinfo {author} {\bibfnamefont {E.~F.}\ \bibnamefont
  {Valeev}}, \ and\ \bibinfo {author} {\bibfnamefont {C.~D.}\ \bibnamefont
  {Sherrill}},\ }\href@noop {} {\bibfield  {journal} {\bibinfo  {journal}
  {Journal of the American Chemical Society}\ }\textbf {\bibinfo {volume}
  {124}},\ \bibinfo {pages} {10887} (\bibinfo {year} {2002})}\BibitemShut
  {NoStop}%
\bibitem [{\citenamefont {Slipchenko}\ and\ \citenamefont
  {Krylov}(2002)}]{slipchenko2002singlet}%
  \BibitemOpen
  \bibfield  {author} {\bibinfo {author} {\bibfnamefont {L.~V.}\ \bibnamefont
  {Slipchenko}}\ and\ \bibinfo {author} {\bibfnamefont {A.~I.}\ \bibnamefont
  {Krylov}},\ }\href@noop {} {\bibfield  {journal} {\bibinfo  {journal} {J.
  Chem. Phys.}\ }\textbf {\bibinfo {volume} {117}},\ \bibinfo {pages} {4694}
  (\bibinfo {year} {2002})}\BibitemShut {NoStop}%
\bibitem [{\citenamefont {Tajti}\ \emph {et~al.}(2004)\citenamefont {Tajti},
  \citenamefont {Szalay}, \citenamefont {Cs{\'a}sz{\'a}r}, \citenamefont
  {K{\'a}llay}, \citenamefont {Gauss}, \citenamefont {Valeev}, \citenamefont
  {Flowers}, \citenamefont {V{\'a}zquez},\ and\ \citenamefont
  {Stanton}}]{tajti2004heat}%
  \BibitemOpen
  \bibfield  {author} {\bibinfo {author} {\bibfnamefont {A.}~\bibnamefont
  {Tajti}}, \bibinfo {author} {\bibfnamefont {P.~G.}\ \bibnamefont {Szalay}},
  \bibinfo {author} {\bibfnamefont {A.~G.}\ \bibnamefont {Cs{\'a}sz{\'a}r}},
  \bibinfo {author} {\bibfnamefont {M.}~\bibnamefont {K{\'a}llay}}, \bibinfo
  {author} {\bibfnamefont {J.}~\bibnamefont {Gauss}}, \bibinfo {author}
  {\bibfnamefont {E.~F.}\ \bibnamefont {Valeev}}, \bibinfo {author}
  {\bibfnamefont {B.~A.}\ \bibnamefont {Flowers}}, \bibinfo {author}
  {\bibfnamefont {J.}~\bibnamefont {V{\'a}zquez}}, \ and\ \bibinfo {author}
  {\bibfnamefont {J.~F.}\ \bibnamefont {Stanton}},\ }\href@noop {} {\bibfield
  {journal} {\bibinfo  {journal} {J. Chem. Phys.}\ }\textbf {\bibinfo {volume}
  {121}},\ \bibinfo {pages} {11599} (\bibinfo {year} {2004})}\BibitemShut
  {NoStop}%
\bibitem [{\citenamefont {Crawford}(2006)}]{crawford2006ab}%
  \BibitemOpen
  \bibfield  {author} {\bibinfo {author} {\bibfnamefont {T.~D.}\ \bibnamefont
  {Crawford}},\ }\href@noop {} {\bibfield  {journal} {\bibinfo  {journal}
  {Theoretical Chemistry Accounts}\ }\textbf {\bibinfo {volume} {115}},\
  \bibinfo {pages} {227} (\bibinfo {year} {2006})}\BibitemShut {NoStop}%
\bibitem [{\citenamefont {Parkhill}, \citenamefont {Lawler},\ and\
  \citenamefont {Head-Gordon}(2009)}]{parkhill2009perfect}%
  \BibitemOpen
  \bibfield  {author} {\bibinfo {author} {\bibfnamefont {J.~A.}\ \bibnamefont
  {Parkhill}}, \bibinfo {author} {\bibfnamefont {K.}~\bibnamefont {Lawler}}, \
  and\ \bibinfo {author} {\bibfnamefont {M.}~\bibnamefont {Head-Gordon}},\
  }\href@noop {} {\bibfield  {journal} {\bibinfo  {journal} {J. Chem. Phys.}\
  }\textbf {\bibinfo {volume} {130}},\ \bibinfo {pages} {084101} (\bibinfo
  {year} {2009})}\BibitemShut {NoStop}%
\bibitem [{\citenamefont {Riplinger}\ and\ \citenamefont
  {Neese}(2013)}]{riplinger2013efficient}%
  \BibitemOpen
  \bibfield  {author} {\bibinfo {author} {\bibfnamefont {C.}~\bibnamefont
  {Riplinger}}\ and\ \bibinfo {author} {\bibfnamefont {F.}~\bibnamefont
  {Neese}},\ }\href@noop {} {\bibfield  {journal} {\bibinfo  {journal} {J.
  Chem. Phys.}\ }\textbf {\bibinfo {volume} {138}},\ \bibinfo {pages} {034106}
  (\bibinfo {year} {2013})}\BibitemShut {NoStop}%
\bibitem [{\citenamefont {Yuwono}, \citenamefont {Magoulas},\ and\
  \citenamefont {Piecuch}(2020)}]{yuwono2020quantum}%
  \BibitemOpen
  \bibfield  {author} {\bibinfo {author} {\bibfnamefont {S.~H.}\ \bibnamefont
  {Yuwono}}, \bibinfo {author} {\bibfnamefont {I.}~\bibnamefont {Magoulas}}, \
  and\ \bibinfo {author} {\bibfnamefont {P.}~\bibnamefont {Piecuch}},\
  }\href@noop {} {\bibfield  {journal} {\bibinfo  {journal} {Science Advances}\
  }\textbf {\bibinfo {volume} {6}},\ \bibinfo {pages} {eaay4058} (\bibinfo
  {year} {2020})}\BibitemShut {NoStop}%
\bibitem [{\citenamefont {Stoll}(1992)}]{stoll1992correlation}%
  \BibitemOpen
  \bibfield  {author} {\bibinfo {author} {\bibfnamefont {H.}~\bibnamefont
  {Stoll}},\ }\href@noop {} {\bibfield  {journal} {\bibinfo  {journal} {Phys.
  Rev. B}\ }\textbf {\bibinfo {volume} {46}},\ \bibinfo {pages} {6700}
  (\bibinfo {year} {1992})}\BibitemShut {NoStop}%
\bibitem [{\citenamefont {Hirata}\ \emph {et~al.}(2004)\citenamefont {Hirata},
  \citenamefont {Podeszwa}, \citenamefont {Tobita},\ and\ \citenamefont
  {Bartlett}}]{hirata2004coupled}%
  \BibitemOpen
  \bibfield  {author} {\bibinfo {author} {\bibfnamefont {S.}~\bibnamefont
  {Hirata}}, \bibinfo {author} {\bibfnamefont {R.}~\bibnamefont {Podeszwa}},
  \bibinfo {author} {\bibfnamefont {M.}~\bibnamefont {Tobita}}, \ and\ \bibinfo
  {author} {\bibfnamefont {R.~J.}\ \bibnamefont {Bartlett}},\ }\href@noop {}
  {\bibfield  {journal} {\bibinfo  {journal} {J. Chem. Phys.}\ }\textbf
  {\bibinfo {volume} {120}},\ \bibinfo {pages} {2581} (\bibinfo {year}
  {2004})}\BibitemShut {NoStop}%
\bibitem [{\citenamefont {Katagiri}(2005)}]{katagiri2005equation}%
  \BibitemOpen
  \bibfield  {author} {\bibinfo {author} {\bibfnamefont {H.}~\bibnamefont
  {Katagiri}},\ }\href@noop {} {\bibfield  {journal} {\bibinfo  {journal} {J.
  Chem. Phys.}\ }\textbf {\bibinfo {volume} {122}},\ \bibinfo {pages} {224901}
  (\bibinfo {year} {2005})}\BibitemShut {NoStop}%
\bibitem [{\citenamefont {Booth}\ \emph {et~al.}(2013)\citenamefont {Booth},
  \citenamefont {Gr{\"u}neis}, \citenamefont {Kresse},\ and\ \citenamefont
  {Alavi}}]{booth2013towards}%
  \BibitemOpen
  \bibfield  {author} {\bibinfo {author} {\bibfnamefont {G.~H.}\ \bibnamefont
  {Booth}}, \bibinfo {author} {\bibfnamefont {A.}~\bibnamefont {Gr{\"u}neis}},
  \bibinfo {author} {\bibfnamefont {G.}~\bibnamefont {Kresse}}, \ and\ \bibinfo
  {author} {\bibfnamefont {A.}~\bibnamefont {Alavi}},\ }\href@noop {}
  {\bibfield  {journal} {\bibinfo  {journal} {Nature}\ }\textbf {\bibinfo
  {volume} {493}},\ \bibinfo {pages} {365} (\bibinfo {year}
  {2013})}\BibitemShut {NoStop}%
\bibitem [{\citenamefont {Degroote}\ \emph {et~al.}(2016)\citenamefont
  {Degroote}, \citenamefont {Henderson}, \citenamefont {Zhao}, \citenamefont
  {Dukelsky},\ and\ \citenamefont {Scuseria}}]{degroote2016polynomial}%
  \BibitemOpen
  \bibfield  {author} {\bibinfo {author} {\bibfnamefont {M.}~\bibnamefont
  {Degroote}}, \bibinfo {author} {\bibfnamefont {T.~M.}\ \bibnamefont
  {Henderson}}, \bibinfo {author} {\bibfnamefont {J.}~\bibnamefont {Zhao}},
  \bibinfo {author} {\bibfnamefont {J.}~\bibnamefont {Dukelsky}}, \ and\
  \bibinfo {author} {\bibfnamefont {G.~E.}\ \bibnamefont {Scuseria}},\
  }\href@noop {} {\bibfield  {journal} {\bibinfo  {journal} {Phys. Rev. B}\
  }\textbf {\bibinfo {volume} {93}},\ \bibinfo {pages} {125124} (\bibinfo
  {year} {2016})}\BibitemShut {NoStop}%
\bibitem [{\citenamefont {McClain}\ \emph {et~al.}(2017)\citenamefont
  {McClain}, \citenamefont {Sun}, \citenamefont {Chan},\ and\ \citenamefont
  {Berkelbach}}]{mcclain2017gaussian}%
  \BibitemOpen
  \bibfield  {author} {\bibinfo {author} {\bibfnamefont {J.}~\bibnamefont
  {McClain}}, \bibinfo {author} {\bibfnamefont {Q.}~\bibnamefont {Sun}},
  \bibinfo {author} {\bibfnamefont {G.~K.-L.}\ \bibnamefont {Chan}}, \ and\
  \bibinfo {author} {\bibfnamefont {T.~C.}\ \bibnamefont {Berkelbach}},\
  }\href@noop {} {\bibfield  {journal} {\bibinfo  {journal} {J. Chem. Theory
  Comput.}\ }\textbf {\bibinfo {volume} {13}},\ \bibinfo {pages} {1209}
  (\bibinfo {year} {2017})}\BibitemShut {NoStop}%
\bibitem [{\citenamefont {Wang}\ and\ \citenamefont
  {Berkelbach}(2020)}]{wang2020excitons}%
  \BibitemOpen
  \bibfield  {author} {\bibinfo {author} {\bibfnamefont {X.}~\bibnamefont
  {Wang}}\ and\ \bibinfo {author} {\bibfnamefont {T.~C.}\ \bibnamefont
  {Berkelbach}},\ }\href@noop {} {\bibfield  {journal} {\bibinfo  {journal} {J.
  Chem. Theory Comput.}\ }\textbf {\bibinfo {volume} {16}},\ \bibinfo {pages}
  {3095} (\bibinfo {year} {2020})}\BibitemShut {NoStop}%
\bibitem [{\citenamefont {Haugland}\ \emph {et~al.}(2020)\citenamefont
  {Haugland}, \citenamefont {Ronca}, \citenamefont {Kj\o{}nstad}, \citenamefont
  {Rubio},\ and\ \citenamefont {Koch}}]{PhysRevX.10.041043}%
  \BibitemOpen
  \bibfield  {author} {\bibinfo {author} {\bibfnamefont {T.~S.}\ \bibnamefont
  {Haugland}}, \bibinfo {author} {\bibfnamefont {E.}~\bibnamefont {Ronca}},
  \bibinfo {author} {\bibfnamefont {E.~F.}\ \bibnamefont {Kj\o{}nstad}},
  \bibinfo {author} {\bibfnamefont {A.}~\bibnamefont {Rubio}}, \ and\ \bibinfo
  {author} {\bibfnamefont {H.}~\bibnamefont {Koch}},\ }\href {\doibase
  10.1103/PhysRevX.10.041043} {\bibfield  {journal} {\bibinfo  {journal} {Phys.
  Rev. X}\ }\textbf {\bibinfo {volume} {10}},\ \bibinfo {pages} {041043}
  (\bibinfo {year} {2020})}\BibitemShut {NoStop}%
\bibitem [{\citenamefont {Farnell}\ and\ \citenamefont
  {Bishop}(2004)}]{farnell2004coupled}%
  \BibitemOpen
  \bibfield  {author} {\bibinfo {author} {\bibfnamefont {D.~J.}\ \bibnamefont
  {Farnell}}\ and\ \bibinfo {author} {\bibfnamefont {R.~F.}\ \bibnamefont
  {Bishop}},\ }in\ \href@noop {} {\emph {\bibinfo {booktitle} {Quantum
  Magnetism}}}\ (\bibinfo  {publisher} {Springer},\ \bibinfo {year} {2004})\
  pp.\ \bibinfo {pages} {307--348}\BibitemShut {NoStop}%
\bibitem [{\citenamefont {Farnell}\ \emph {et~al.}(2018)\citenamefont
  {Farnell}, \citenamefont {G{\"o}tze}, \citenamefont {Schulenburg},
  \citenamefont {Zinke}, \citenamefont {Bishop},\ and\ \citenamefont
  {Li}}]{farnell2018interplay}%
  \BibitemOpen
  \bibfield  {author} {\bibinfo {author} {\bibfnamefont {D.}~\bibnamefont
  {Farnell}}, \bibinfo {author} {\bibfnamefont {O.}~\bibnamefont {G{\"o}tze}},
  \bibinfo {author} {\bibfnamefont {J.}~\bibnamefont {Schulenburg}}, \bibinfo
  {author} {\bibfnamefont {R.}~\bibnamefont {Zinke}}, \bibinfo {author}
  {\bibfnamefont {R.}~\bibnamefont {Bishop}}, \ and\ \bibinfo {author}
  {\bibfnamefont {P.}~\bibnamefont {Li}},\ }\href@noop {} {\bibfield  {journal}
  {\bibinfo  {journal} {Physical Review B}\ }\textbf {\bibinfo {volume} {98}},\
  \bibinfo {pages} {224402} (\bibinfo {year} {2018})}\BibitemShut {NoStop}%
\bibitem [{\citenamefont {Bishop}\ \emph {et~al.}(2019)\citenamefont {Bishop},
  \citenamefont {Li}, \citenamefont {G{\"o}tze},\ and\ \citenamefont
  {Richter}}]{bishop2019frustrated}%
  \BibitemOpen
  \bibfield  {author} {\bibinfo {author} {\bibfnamefont {R.}~\bibnamefont
  {Bishop}}, \bibinfo {author} {\bibfnamefont {P.~H.}\ \bibnamefont {Li}},
  \bibinfo {author} {\bibfnamefont {O.}~\bibnamefont {G{\"o}tze}}, \ and\
  \bibinfo {author} {\bibfnamefont {J.}~\bibnamefont {Richter}},\ }\href@noop
  {} {\bibfield  {journal} {\bibinfo  {journal} {Physical Review B}\ }\textbf
  {\bibinfo {volume} {100}},\ \bibinfo {pages} {024401} (\bibinfo {year}
  {2019})}\BibitemShut {NoStop}%
\bibitem [{\citenamefont {Brandow}(1967)}]{brandow67_771}%
  \BibitemOpen
  \bibfield  {author} {\bibinfo {author} {\bibfnamefont {B.~H.}\ \bibnamefont
  {Brandow}},\ }\href {\doibase 10.1103/RevModPhys.39.771} {\bibfield
  {journal} {\bibinfo  {journal} {Rev. Mod. Phys.}\ }\textbf {\bibinfo {volume}
  {39}},\ \bibinfo {pages} {771} (\bibinfo {year} {1967})}\BibitemShut
  {NoStop}%
\bibitem [{\citenamefont {Lindgren}\ and\ \citenamefont
  {Morrison}(2012)}]{lindgren12}%
  \BibitemOpen
  \bibfield  {author} {\bibinfo {author} {\bibfnamefont {I.}~\bibnamefont
  {Lindgren}}\ and\ \bibinfo {author} {\bibfnamefont {J.}~\bibnamefont
  {Morrison}},\ }\href {https://books.google.com/books?id=L43\_CAAAQBAJ} {\emph
  {\bibinfo {title} {Atomic Many-Body Theory}}},\ Springer Series on Atomic,
  Optical, and Plasma Physics\ (\bibinfo  {publisher} {Springer Berlin
  Heidelberg},\ \bibinfo {year} {2012})\BibitemShut {NoStop}%
\bibitem [{\citenamefont {Kowalski}(2018)}]{safkk}%
  \BibitemOpen
  \bibfield  {author} {\bibinfo {author} {\bibfnamefont {K.}~\bibnamefont
  {Kowalski}},\ }\href {\doibase 10.1063/1.5010693} {\bibfield  {journal}
  {\bibinfo  {journal} {J. Chem. Phys.}\ }\textbf {\bibinfo {volume} {148}},\
  \bibinfo {pages} {094104} (\bibinfo {year} {2018})}\BibitemShut {NoStop}%
\bibitem [{\citenamefont {Bauman}\ \emph {et~al.}(2019)\citenamefont {Bauman},
  \citenamefont {Bylaska}, \citenamefont {Krishnamoorthy}, \citenamefont {Low},
  \citenamefont {Wiebe}, \citenamefont {Granade}, \citenamefont {Roetteler},
  \citenamefont {Troyer},\ and\ \citenamefont
  {Kowalski}}]{bauman2019downfolding}%
  \BibitemOpen
  \bibfield  {author} {\bibinfo {author} {\bibfnamefont {N.~P.}\ \bibnamefont
  {Bauman}}, \bibinfo {author} {\bibfnamefont {E.~J.}\ \bibnamefont {Bylaska}},
  \bibinfo {author} {\bibfnamefont {S.}~\bibnamefont {Krishnamoorthy}},
  \bibinfo {author} {\bibfnamefont {G.~H.}\ \bibnamefont {Low}}, \bibinfo
  {author} {\bibfnamefont {N.}~\bibnamefont {Wiebe}}, \bibinfo {author}
  {\bibfnamefont {C.~E.}\ \bibnamefont {Granade}}, \bibinfo {author}
  {\bibfnamefont {M.}~\bibnamefont {Roetteler}}, \bibinfo {author}
  {\bibfnamefont {M.}~\bibnamefont {Troyer}}, \ and\ \bibinfo {author}
  {\bibfnamefont {K.}~\bibnamefont {Kowalski}},\ }\href@noop {} {\bibfield
  {journal} {\bibinfo  {journal} {J. Chem. Phys.}\ }\textbf {\bibinfo {volume}
  {151}},\ \bibinfo {pages} {014107} (\bibinfo {year} {2019})}\BibitemShut
  {NoStop}%
\bibitem [{\citenamefont {Bauman}, \citenamefont {Low},\ and\ \citenamefont
  {Kowalski}(2019)}]{bauman2019quantumex}%
  \BibitemOpen
  \bibfield  {author} {\bibinfo {author} {\bibfnamefont {N.~P.}\ \bibnamefont
  {Bauman}}, \bibinfo {author} {\bibfnamefont {G.~H.}\ \bibnamefont {Low}}, \
  and\ \bibinfo {author} {\bibfnamefont {K.}~\bibnamefont {Kowalski}},\
  }\href@noop {} {\bibfield  {journal} {\bibinfo  {journal} {J. Chem. Phys.}\
  }\textbf {\bibinfo {volume} {151}},\ \bibinfo {pages} {234114} (\bibinfo
  {year} {2019})}\BibitemShut {NoStop}%
\bibitem [{\citenamefont {Kowalski}\ and\ \citenamefont
  {Bauman}(2020)}]{downfolding2020t}%
  \BibitemOpen
  \bibfield  {author} {\bibinfo {author} {\bibfnamefont {K.}~\bibnamefont
  {Kowalski}}\ and\ \bibinfo {author} {\bibfnamefont {N.~P.}\ \bibnamefont
  {Bauman}},\ }\href {\doibase 10.1063/5.0008436} {\bibfield  {journal}
  {\bibinfo  {journal} {J. Chem. Phys.}\ }\textbf {\bibinfo {volume} {152}},\
  \bibinfo {pages} {244127} (\bibinfo {year} {2020})}\BibitemShut {NoStop}%
\bibitem [{\citenamefont {Kowalski}(2021)}]{kowalski2021dimensionality}%
  \BibitemOpen
  \bibfield  {author} {\bibinfo {author} {\bibfnamefont {K.}~\bibnamefont
  {Kowalski}},\ }\href {\doibase 10.1103/PhysRevA.104.032804} {\bibfield
  {journal} {\bibinfo  {journal} {Phys. Rev. A}\ }\textbf {\bibinfo {volume}
  {104}},\ \bibinfo {pages} {032804} (\bibinfo {year} {2021})}\BibitemShut
  {NoStop}%
\bibitem [{\citenamefont {Bauman}\ and\ \citenamefont
  {Kowalski}(2022{\natexlab{a}})}]{bauman2022coupled}%
  \BibitemOpen
  \bibfield  {author} {\bibinfo {author} {\bibfnamefont {N.~P.}\ \bibnamefont
  {Bauman}}\ and\ \bibinfo {author} {\bibfnamefont {K.}~\bibnamefont
  {Kowalski}},\ }\href@noop {} {\bibfield  {journal} {\bibinfo  {journal}
  {Materials Theory}\ }\textbf {\bibinfo {volume} {6}},\ \bibinfo {pages} {1}
  (\bibinfo {year} {2022}{\natexlab{a}})}\BibitemShut {NoStop}%
\bibitem [{\citenamefont {Bauman}\ and\ \citenamefont
  {Kowalski}(2022{\natexlab{b}})}]{bauman2022coupled2c}%
  \BibitemOpen
  \bibfield  {author} {\bibinfo {author} {\bibfnamefont {N.~P.}\ \bibnamefont
  {Bauman}}\ and\ \bibinfo {author} {\bibfnamefont {K.}~\bibnamefont
  {Kowalski}},\ }\href@noop {} {\bibfield  {journal} {\bibinfo  {journal} {The
  Journal of Chemical Physics}\ }\textbf {\bibinfo {volume} {156}},\ \bibinfo
  {pages} {094106} (\bibinfo {year} {2022}{\natexlab{b}})}\BibitemShut
  {NoStop}%
\bibitem [{\citenamefont {He}, \citenamefont {Li},\ and\ \citenamefont
  {Evangelista}(2022)}]{he2022second}%
  \BibitemOpen
  \bibfield  {author} {\bibinfo {author} {\bibfnamefont {N.}~\bibnamefont
  {He}}, \bibinfo {author} {\bibfnamefont {C.}~\bibnamefont {Li}}, \ and\
  \bibinfo {author} {\bibfnamefont {F.~A.}\ \bibnamefont {Evangelista}},\
  }\href@noop {} {\bibfield  {journal} {\bibinfo  {journal} {Journal of
  Chemical Theory and Computation}\ }\textbf {\bibinfo {volume} {18}},\
  \bibinfo {pages} {1527} (\bibinfo {year} {2022})}\BibitemShut {NoStop}%
\bibitem [{\citenamefont {Luis}\ and\ \citenamefont
  {Pe{\v{r}}ina}(1996)}]{luis1996optimum}%
  \BibitemOpen
  \bibfield  {author} {\bibinfo {author} {\bibfnamefont {A.}~\bibnamefont
  {Luis}}\ and\ \bibinfo {author} {\bibfnamefont {J.}~\bibnamefont
  {Pe{\v{r}}ina}},\ }\href@noop {} {\bibfield  {journal} {\bibinfo  {journal}
  {Phys. Rev. A}\ }\textbf {\bibinfo {volume} {54}},\ \bibinfo {pages} {4564}
  (\bibinfo {year} {1996})}\BibitemShut {NoStop}%
\bibitem [{\citenamefont {Cleve}\ \emph {et~al.}(1998)\citenamefont {Cleve},
  \citenamefont {Ekert}, \citenamefont {Macchiavello},\ and\ \citenamefont
  {Mosca}}]{cleve1998quantum}%
  \BibitemOpen
  \bibfield  {author} {\bibinfo {author} {\bibfnamefont {R.}~\bibnamefont
  {Cleve}}, \bibinfo {author} {\bibfnamefont {A.}~\bibnamefont {Ekert}},
  \bibinfo {author} {\bibfnamefont {C.}~\bibnamefont {Macchiavello}}, \ and\
  \bibinfo {author} {\bibfnamefont {M.}~\bibnamefont {Mosca}},\ }\href@noop {}
  {\bibfield  {journal} {\bibinfo  {journal} {Proc. R. Soc. Lond. A}\ }\textbf
  {\bibinfo {volume} {454}},\ \bibinfo {pages} {339} (\bibinfo {year}
  {1998})}\BibitemShut {NoStop}%
\bibitem [{\citenamefont {Berry}\ \emph {et~al.}(2007)\citenamefont {Berry},
  \citenamefont {Ahokas}, \citenamefont {Cleve},\ and\ \citenamefont
  {Sanders}}]{berry2007efficient}%
  \BibitemOpen
  \bibfield  {author} {\bibinfo {author} {\bibfnamefont {D.~W.}\ \bibnamefont
  {Berry}}, \bibinfo {author} {\bibfnamefont {G.}~\bibnamefont {Ahokas}},
  \bibinfo {author} {\bibfnamefont {R.}~\bibnamefont {Cleve}}, \ and\ \bibinfo
  {author} {\bibfnamefont {B.~C.}\ \bibnamefont {Sanders}},\ }\href@noop {}
  {\bibfield  {journal} {\bibinfo  {journal} {Comm. Math. Phys.}\ }\textbf
  {\bibinfo {volume} {270}},\ \bibinfo {pages} {359} (\bibinfo {year}
  {2007})}\BibitemShut {NoStop}%
\bibitem [{\citenamefont {Childs}(2010)}]{childs2010relationship}%
  \BibitemOpen
  \bibfield  {author} {\bibinfo {author} {\bibfnamefont {A.~M.}\ \bibnamefont
  {Childs}},\ }\href@noop {} {\bibfield  {journal} {\bibinfo  {journal} {Comm.
  Math. Phys.}\ }\textbf {\bibinfo {volume} {294}},\ \bibinfo {pages} {581}
  (\bibinfo {year} {2010})}\BibitemShut {NoStop}%
\bibitem [{\citenamefont {Wecker}, \citenamefont {Hastings},\ and\
  \citenamefont {Troyer}(2015)}]{wecker2015progress}%
  \BibitemOpen
  \bibfield  {author} {\bibinfo {author} {\bibfnamefont {D.}~\bibnamefont
  {Wecker}}, \bibinfo {author} {\bibfnamefont {M.~B.}\ \bibnamefont
  {Hastings}}, \ and\ \bibinfo {author} {\bibfnamefont {M.}~\bibnamefont
  {Troyer}},\ }\href@noop {} {\bibfield  {journal} {\bibinfo  {journal} {Phys.
  Rev. A}\ }\textbf {\bibinfo {volume} {92}},\ \bibinfo {pages} {042303}
  (\bibinfo {year} {2015})}\BibitemShut {NoStop}%
\bibitem [{\citenamefont {H\"aner}\ \emph {et~al.}(2016)\citenamefont
  {H\"aner}, \citenamefont {Steiger}, \citenamefont {Smelyanskiy},\ and\
  \citenamefont {Troyer}}]{haner2016high}%
  \BibitemOpen
  \bibfield  {author} {\bibinfo {author} {\bibfnamefont {T.}~\bibnamefont
  {H\"aner}}, \bibinfo {author} {\bibfnamefont {D.~S.}\ \bibnamefont
  {Steiger}}, \bibinfo {author} {\bibfnamefont {M.}~\bibnamefont
  {Smelyanskiy}}, \ and\ \bibinfo {author} {\bibfnamefont {M.}~\bibnamefont
  {Troyer}},\ }in\ \href {\doibase 10.1109/SC.2016.73} {\emph {\bibinfo
  {booktitle} {SC '16: Proceedings of the International Conference for High
  Performance Computing, Networking, Storage and Analysis}}}\ (\bibinfo {year}
  {2016})\ pp.\ \bibinfo {pages} {866--874}\BibitemShut {NoStop}%
\bibitem [{\citenamefont {Poulin}\ \emph {et~al.}(2017)\citenamefont {Poulin},
  \citenamefont {Kitaev}, \citenamefont {Steiger}, \citenamefont {Hastings},\
  and\ \citenamefont {Troyer}}]{poulin2017fast}%
  \BibitemOpen
  \bibfield  {author} {\bibinfo {author} {\bibfnamefont {D.}~\bibnamefont
  {Poulin}}, \bibinfo {author} {\bibfnamefont {A.}~\bibnamefont {Kitaev}},
  \bibinfo {author} {\bibfnamefont {D.~S.}\ \bibnamefont {Steiger}}, \bibinfo
  {author} {\bibfnamefont {M.~B.}\ \bibnamefont {Hastings}}, \ and\ \bibinfo
  {author} {\bibfnamefont {M.}~\bibnamefont {Troyer}},\ }\href@noop {}
  {\bibfield  {journal} {\bibinfo  {journal} {arXiv preprint arXiv:1711.11025}\
  } (\bibinfo {year} {2017})}\BibitemShut {NoStop}%
\bibitem [{\citenamefont {Peruzzo}\ \emph {et~al.}(2014)\citenamefont
  {Peruzzo}, \citenamefont {McClean}, \citenamefont {Shadbolt}, \citenamefont
  {Yung}, \citenamefont {Zhou}, \citenamefont {Love}, \citenamefont
  {Aspuru-Guzik},\ and\ \citenamefont {O'brien}}]{peruzzo2014variational}%
  \BibitemOpen
  \bibfield  {author} {\bibinfo {author} {\bibfnamefont {A.}~\bibnamefont
  {Peruzzo}}, \bibinfo {author} {\bibfnamefont {J.}~\bibnamefont {McClean}},
  \bibinfo {author} {\bibfnamefont {P.}~\bibnamefont {Shadbolt}}, \bibinfo
  {author} {\bibfnamefont {M.-H.}\ \bibnamefont {Yung}}, \bibinfo {author}
  {\bibfnamefont {X.-Q.}\ \bibnamefont {Zhou}}, \bibinfo {author}
  {\bibfnamefont {P.~J.}\ \bibnamefont {Love}}, \bibinfo {author}
  {\bibfnamefont {A.}~\bibnamefont {Aspuru-Guzik}}, \ and\ \bibinfo {author}
  {\bibfnamefont {J.~L.}\ \bibnamefont {O'brien}},\ }\href@noop {} {\bibfield
  {journal} {\bibinfo  {journal} {Nat. Commun.}\ }\textbf {\bibinfo {volume}
  {5}},\ \bibinfo {pages} {4213} (\bibinfo {year} {2014})}\BibitemShut
  {NoStop}%
\bibitem [{\citenamefont {McClean}\ \emph {et~al.}(2016)\citenamefont
  {McClean}, \citenamefont {Romero}, \citenamefont {Babbush},\ and\
  \citenamefont {Aspuru-Guzik}}]{mcclean2016theory}%
  \BibitemOpen
  \bibfield  {author} {\bibinfo {author} {\bibfnamefont {J.~R.}\ \bibnamefont
  {McClean}}, \bibinfo {author} {\bibfnamefont {J.}~\bibnamefont {Romero}},
  \bibinfo {author} {\bibfnamefont {R.}~\bibnamefont {Babbush}}, \ and\
  \bibinfo {author} {\bibfnamefont {A.}~\bibnamefont {Aspuru-Guzik}},\
  }\href@noop {} {\bibfield  {journal} {\bibinfo  {journal} {New J. Phys.}\
  }\textbf {\bibinfo {volume} {18}},\ \bibinfo {pages} {023023} (\bibinfo
  {year} {2016})}\BibitemShut {NoStop}%
\bibitem [{\citenamefont {Romero}\ \emph {et~al.}(2018)\citenamefont {Romero},
  \citenamefont {Babbush}, \citenamefont {McClean}, \citenamefont {Hempel},
  \citenamefont {Love},\ and\ \citenamefont
  {Aspuru-Guzik}}]{romero2018strategies}%
  \BibitemOpen
  \bibfield  {author} {\bibinfo {author} {\bibfnamefont {J.}~\bibnamefont
  {Romero}}, \bibinfo {author} {\bibfnamefont {R.}~\bibnamefont {Babbush}},
  \bibinfo {author} {\bibfnamefont {J.~R.}\ \bibnamefont {McClean}}, \bibinfo
  {author} {\bibfnamefont {C.}~\bibnamefont {Hempel}}, \bibinfo {author}
  {\bibfnamefont {P.~J.}\ \bibnamefont {Love}}, \ and\ \bibinfo {author}
  {\bibfnamefont {A.}~\bibnamefont {Aspuru-Guzik}},\ }\href@noop {} {\bibfield
  {journal} {\bibinfo  {journal} {Quantum Sci. Technol.}\ }\textbf {\bibinfo
  {volume} {4}},\ \bibinfo {pages} {014008} (\bibinfo {year}
  {2018})}\BibitemShut {NoStop}%
\bibitem [{\citenamefont {Shen}\ \emph {et~al.}(2017)\citenamefont {Shen},
  \citenamefont {Zhang}, \citenamefont {Zhang}, \citenamefont {Zhang},
  \citenamefont {Yung},\ and\ \citenamefont {Kim}}]{PhysRevA.95.020501}%
  \BibitemOpen
  \bibfield  {author} {\bibinfo {author} {\bibfnamefont {Y.}~\bibnamefont
  {Shen}}, \bibinfo {author} {\bibfnamefont {X.}~\bibnamefont {Zhang}},
  \bibinfo {author} {\bibfnamefont {S.}~\bibnamefont {Zhang}}, \bibinfo
  {author} {\bibfnamefont {J.-N.}\ \bibnamefont {Zhang}}, \bibinfo {author}
  {\bibfnamefont {M.-H.}\ \bibnamefont {Yung}}, \ and\ \bibinfo {author}
  {\bibfnamefont {K.}~\bibnamefont {Kim}},\ }\href {\doibase
  10.1103/PhysRevA.95.020501} {\bibfield  {journal} {\bibinfo  {journal} {Phys.
  Rev. A}\ }\textbf {\bibinfo {volume} {95}},\ \bibinfo {pages} {020501}
  (\bibinfo {year} {2017})}\BibitemShut {NoStop}%
\bibitem [{\citenamefont {Kandala}\ \emph {et~al.}(2017)\citenamefont
  {Kandala}, \citenamefont {Mezzacapo}, \citenamefont {Temme}, \citenamefont
  {Takita}, \citenamefont {Brink}, \citenamefont {Chow},\ and\ \citenamefont
  {Gambetta}}]{Kandala2017}%
  \BibitemOpen
  \bibfield  {author} {\bibinfo {author} {\bibfnamefont {A.}~\bibnamefont
  {Kandala}}, \bibinfo {author} {\bibfnamefont {A.}~\bibnamefont {Mezzacapo}},
  \bibinfo {author} {\bibfnamefont {K.}~\bibnamefont {Temme}}, \bibinfo
  {author} {\bibfnamefont {M.}~\bibnamefont {Takita}}, \bibinfo {author}
  {\bibfnamefont {M.}~\bibnamefont {Brink}}, \bibinfo {author} {\bibfnamefont
  {J.~M.}\ \bibnamefont {Chow}}, \ and\ \bibinfo {author} {\bibfnamefont
  {J.~M.}\ \bibnamefont {Gambetta}},\ }\href@noop {} {\bibfield  {journal}
  {\bibinfo  {journal} {Nature}\ }\textbf {\bibinfo {volume} {549}},\ \bibinfo
  {pages} {242} (\bibinfo {year} {2017})}\BibitemShut {NoStop}%
\bibitem [{\citenamefont {Kandala}\ \emph {et~al.}(2019)\citenamefont
  {Kandala}, \citenamefont {Temme}, \citenamefont {Corcoles}, \citenamefont
  {Mezzacapo}, \citenamefont {Chow},\ and\ \citenamefont
  {Gambetta}}]{kandala2018extending}%
  \BibitemOpen
  \bibfield  {author} {\bibinfo {author} {\bibfnamefont {A.}~\bibnamefont
  {Kandala}}, \bibinfo {author} {\bibfnamefont {K.}~\bibnamefont {Temme}},
  \bibinfo {author} {\bibfnamefont {A.~D.}\ \bibnamefont {Corcoles}}, \bibinfo
  {author} {\bibfnamefont {A.}~\bibnamefont {Mezzacapo}}, \bibinfo {author}
  {\bibfnamefont {J.~M.}\ \bibnamefont {Chow}}, \ and\ \bibinfo {author}
  {\bibfnamefont {J.~M.}\ \bibnamefont {Gambetta}},\ }\href@noop {} {\bibfield
  {journal} {\bibinfo  {journal} {Nature}\ }\textbf {\bibinfo {volume} {567}},\
  \bibinfo {pages} {491} (\bibinfo {year} {2019})}\BibitemShut {NoStop}%
\bibitem [{\citenamefont {Colless}\ \emph {et~al.}(2018)\citenamefont
  {Colless}, \citenamefont {Ramasesh}, \citenamefont {Dahlen}, \citenamefont
  {Blok}, \citenamefont {Kimchi-Schwartz}, \citenamefont {McClean},
  \citenamefont {Carter}, \citenamefont {de~Jong},\ and\ \citenamefont
  {Siddiqi}}]{PhysRevX.8.011021}%
  \BibitemOpen
  \bibfield  {author} {\bibinfo {author} {\bibfnamefont {J.~I.}\ \bibnamefont
  {Colless}}, \bibinfo {author} {\bibfnamefont {V.~V.}\ \bibnamefont
  {Ramasesh}}, \bibinfo {author} {\bibfnamefont {D.}~\bibnamefont {Dahlen}},
  \bibinfo {author} {\bibfnamefont {M.~S.}\ \bibnamefont {Blok}}, \bibinfo
  {author} {\bibfnamefont {M.~E.}\ \bibnamefont {Kimchi-Schwartz}}, \bibinfo
  {author} {\bibfnamefont {J.~R.}\ \bibnamefont {McClean}}, \bibinfo {author}
  {\bibfnamefont {J.}~\bibnamefont {Carter}}, \bibinfo {author} {\bibfnamefont
  {W.~A.}\ \bibnamefont {de~Jong}}, \ and\ \bibinfo {author} {\bibfnamefont
  {I.}~\bibnamefont {Siddiqi}},\ }\href {\doibase 10.1103/PhysRevX.8.011021}
  {\bibfield  {journal} {\bibinfo  {journal} {Phys. Rev. X}\ }\textbf {\bibinfo
  {volume} {8}},\ \bibinfo {pages} {011021} (\bibinfo {year}
  {2018})}\BibitemShut {NoStop}%
\bibitem [{\citenamefont {Huggins}\ \emph {et~al.}(2020)\citenamefont
  {Huggins}, \citenamefont {Lee}, \citenamefont {Baek}, \citenamefont
  {O'Gorman},\ and\ \citenamefont {Whaley}}]{huggins2020non}%
  \BibitemOpen
  \bibfield  {author} {\bibinfo {author} {\bibfnamefont {W.~J.}\ \bibnamefont
  {Huggins}}, \bibinfo {author} {\bibfnamefont {J.}~\bibnamefont {Lee}},
  \bibinfo {author} {\bibfnamefont {U.}~\bibnamefont {Baek}}, \bibinfo {author}
  {\bibfnamefont {B.}~\bibnamefont {O'Gorman}}, \ and\ \bibinfo {author}
  {\bibfnamefont {K.~B.}\ \bibnamefont {Whaley}},\ }\href@noop {} {\bibfield
  {journal} {\bibinfo  {journal} {New J. Phys.}\ }\textbf {\bibinfo {volume}
  {22}},\ \bibinfo {pages} {073009} (\bibinfo {year} {2020})}\BibitemShut
  {NoStop}%
\bibitem [{\citenamefont {Ryabinkin}\ \emph {et~al.}(2018)\citenamefont
  {Ryabinkin}, \citenamefont {Yen}, \citenamefont {Genin},\ and\ \citenamefont
  {Izmaylov}}]{ryabinkin2018qubit}%
  \BibitemOpen
  \bibfield  {author} {\bibinfo {author} {\bibfnamefont {I.~G.}\ \bibnamefont
  {Ryabinkin}}, \bibinfo {author} {\bibfnamefont {T.-C.}\ \bibnamefont {Yen}},
  \bibinfo {author} {\bibfnamefont {S.~N.}\ \bibnamefont {Genin}}, \ and\
  \bibinfo {author} {\bibfnamefont {A.~F.}\ \bibnamefont {Izmaylov}},\
  }\href@noop {} {\bibfield  {journal} {\bibinfo  {journal} {J. Chem. Theory
  Comput.}\ }\textbf {\bibinfo {volume} {14}},\ \bibinfo {pages} {6317}
  (\bibinfo {year} {2018})}\BibitemShut {NoStop}%
\bibitem [{\citenamefont {Cao}\ \emph {et~al.}(2019)\citenamefont {Cao},
  \citenamefont {Romero}, \citenamefont {Olson}, \citenamefont {Degroote},
  \citenamefont {Johnson}, \citenamefont {Kieferov{\'a}}, \citenamefont
  {Kivlichan}, \citenamefont {Menke}, \citenamefont {Peropadre}, \citenamefont
  {Sawaya} \emph {et~al.}}]{cao2019quantum}%
  \BibitemOpen
  \bibfield  {author} {\bibinfo {author} {\bibfnamefont {Y.}~\bibnamefont
  {Cao}}, \bibinfo {author} {\bibfnamefont {J.}~\bibnamefont {Romero}},
  \bibinfo {author} {\bibfnamefont {J.~P.}\ \bibnamefont {Olson}}, \bibinfo
  {author} {\bibfnamefont {M.}~\bibnamefont {Degroote}}, \bibinfo {author}
  {\bibfnamefont {P.~D.}\ \bibnamefont {Johnson}}, \bibinfo {author}
  {\bibfnamefont {M.}~\bibnamefont {Kieferov{\'a}}}, \bibinfo {author}
  {\bibfnamefont {I.~D.}\ \bibnamefont {Kivlichan}}, \bibinfo {author}
  {\bibfnamefont {T.}~\bibnamefont {Menke}}, \bibinfo {author} {\bibfnamefont
  {B.}~\bibnamefont {Peropadre}}, \bibinfo {author} {\bibfnamefont {N.~P.}\
  \bibnamefont {Sawaya}},  \emph {et~al.},\ }\href@noop {} {\bibfield
  {journal} {\bibinfo  {journal} {Chem. Rev.}\ }\textbf {\bibinfo {volume}
  {119}},\ \bibinfo {pages} {10856} (\bibinfo {year} {2019})}\BibitemShut
  {NoStop}%
\bibitem [{\citenamefont {Ryabinkin}\ \emph {et~al.}(2020)\citenamefont
  {Ryabinkin}, \citenamefont {Lang}, \citenamefont {Genin},\ and\ \citenamefont
  {Izmaylov}}]{ryabinkin2020iterative}%
  \BibitemOpen
  \bibfield  {author} {\bibinfo {author} {\bibfnamefont {I.~G.}\ \bibnamefont
  {Ryabinkin}}, \bibinfo {author} {\bibfnamefont {R.~A.}\ \bibnamefont {Lang}},
  \bibinfo {author} {\bibfnamefont {S.~N.}\ \bibnamefont {Genin}}, \ and\
  \bibinfo {author} {\bibfnamefont {A.~F.}\ \bibnamefont {Izmaylov}},\
  }\href@noop {} {\bibfield  {journal} {\bibinfo  {journal} {J. Chem. Theory
  Comput.}\ }\textbf {\bibinfo {volume} {16}},\ \bibinfo {pages} {1055}
  (\bibinfo {year} {2020})}\BibitemShut {NoStop}%
\bibitem [{\citenamefont {Izmaylov}\ \emph {et~al.}(2019)\citenamefont
  {Izmaylov}, \citenamefont {Yen}, \citenamefont {Lang},\ and\ \citenamefont
  {Verteletskyi}}]{izmaylov2019unitary}%
  \BibitemOpen
  \bibfield  {author} {\bibinfo {author} {\bibfnamefont {A.~F.}\ \bibnamefont
  {Izmaylov}}, \bibinfo {author} {\bibfnamefont {T.-C.}\ \bibnamefont {Yen}},
  \bibinfo {author} {\bibfnamefont {R.~A.}\ \bibnamefont {Lang}}, \ and\
  \bibinfo {author} {\bibfnamefont {V.}~\bibnamefont {Verteletskyi}},\
  }\href@noop {} {\bibfield  {journal} {\bibinfo  {journal} {J. Chem. Theory
  Comput.}\ }\textbf {\bibinfo {volume} {16}},\ \bibinfo {pages} {190}
  (\bibinfo {year} {2019})}\BibitemShut {NoStop}%
\bibitem [{\citenamefont {Lang}, \citenamefont {Ryabinkin},\ and\ \citenamefont
  {Izmaylov}(2021)}]{lang2020unitary}%
  \BibitemOpen
  \bibfield  {author} {\bibinfo {author} {\bibfnamefont {R.~A.}\ \bibnamefont
  {Lang}}, \bibinfo {author} {\bibfnamefont {I.~G.}\ \bibnamefont {Ryabinkin}},
  \ and\ \bibinfo {author} {\bibfnamefont {A.~F.}\ \bibnamefont {Izmaylov}},\
  }\href@noop {} {\bibfield  {journal} {\bibinfo  {journal} {JJ. Chem. Theory
  Comput.}\ }\textbf {\bibinfo {volume} {17}},\ \bibinfo {pages} {66} (\bibinfo
  {year} {2021})}\BibitemShut {NoStop}%
\bibitem [{\citenamefont {Grimsley}\ \emph
  {et~al.}(2019{\natexlab{a}})\citenamefont {Grimsley}, \citenamefont
  {Economou}, \citenamefont {Barnes},\ and\ \citenamefont
  {Mayhall}}]{grimsley2019adaptive}%
  \BibitemOpen
  \bibfield  {author} {\bibinfo {author} {\bibfnamefont {H.~R.}\ \bibnamefont
  {Grimsley}}, \bibinfo {author} {\bibfnamefont {S.~E.}\ \bibnamefont
  {Economou}}, \bibinfo {author} {\bibfnamefont {E.}~\bibnamefont {Barnes}}, \
  and\ \bibinfo {author} {\bibfnamefont {N.~J.}\ \bibnamefont {Mayhall}},\
  }\href@noop {} {\bibfield  {journal} {\bibinfo  {journal} {Nat. Commun.}\
  }\textbf {\bibinfo {volume} {10}},\ \bibinfo {pages} {1} (\bibinfo {year}
  {2019}{\natexlab{a}})}\BibitemShut {NoStop}%
\bibitem [{\citenamefont {Grimsley}\ \emph
  {et~al.}(2019{\natexlab{b}})\citenamefont {Grimsley}, \citenamefont
  {Claudino}, \citenamefont {Economou}, \citenamefont {Barnes},\ and\
  \citenamefont {Mayhall}}]{grimsley2019trotterized}%
  \BibitemOpen
  \bibfield  {author} {\bibinfo {author} {\bibfnamefont {H.~R.}\ \bibnamefont
  {Grimsley}}, \bibinfo {author} {\bibfnamefont {D.}~\bibnamefont {Claudino}},
  \bibinfo {author} {\bibfnamefont {S.~E.}\ \bibnamefont {Economou}}, \bibinfo
  {author} {\bibfnamefont {E.}~\bibnamefont {Barnes}}, \ and\ \bibinfo {author}
  {\bibfnamefont {N.~J.}\ \bibnamefont {Mayhall}},\ }\href@noop {} {\bibfield
  {journal} {\bibinfo  {journal} {J. Chem. Theory Comput.}\ }\textbf {\bibinfo
  {volume} {16}},\ \bibinfo {pages} {1} (\bibinfo {year}
  {2019}{\natexlab{b}})}\BibitemShut {NoStop}%
\bibitem [{\citenamefont {Cerezo}\ \emph {et~al.}(2021)\citenamefont {Cerezo},
  \citenamefont {Arrasmith}, \citenamefont {Babbush}, \citenamefont {Benjamin},
  \citenamefont {Endo}, \citenamefont {Fujii}, \citenamefont {McClean},
  \citenamefont {Mitarai}, \citenamefont {Yuan}, \citenamefont {Cincio} \emph
  {et~al.}}]{cerezo2021variational}%
  \BibitemOpen
  \bibfield  {author} {\bibinfo {author} {\bibfnamefont {M.}~\bibnamefont
  {Cerezo}}, \bibinfo {author} {\bibfnamefont {A.}~\bibnamefont {Arrasmith}},
  \bibinfo {author} {\bibfnamefont {R.}~\bibnamefont {Babbush}}, \bibinfo
  {author} {\bibfnamefont {S.~C.}\ \bibnamefont {Benjamin}}, \bibinfo {author}
  {\bibfnamefont {S.}~\bibnamefont {Endo}}, \bibinfo {author} {\bibfnamefont
  {K.}~\bibnamefont {Fujii}}, \bibinfo {author} {\bibfnamefont {J.~R.}\
  \bibnamefont {McClean}}, \bibinfo {author} {\bibfnamefont {K.}~\bibnamefont
  {Mitarai}}, \bibinfo {author} {\bibfnamefont {X.}~\bibnamefont {Yuan}},
  \bibinfo {author} {\bibfnamefont {L.}~\bibnamefont {Cincio}},  \emph
  {et~al.},\ }\href@noop {} {\bibfield  {journal} {\bibinfo  {journal} {Nature
  Reviews Physics}\ }\textbf {\bibinfo {volume} {3}},\ \bibinfo {pages} {625}
  (\bibinfo {year} {2021})}\BibitemShut {NoStop}%
\bibitem [{\citenamefont {McArdle}\ \emph {et~al.}(2020)\citenamefont
  {McArdle}, \citenamefont {Endo}, \citenamefont {Aspuru-Guzik}, \citenamefont
  {Benjamin},\ and\ \citenamefont {Yuan}}]{mcardle2020quantum}%
  \BibitemOpen
  \bibfield  {author} {\bibinfo {author} {\bibfnamefont {S.}~\bibnamefont
  {McArdle}}, \bibinfo {author} {\bibfnamefont {S.}~\bibnamefont {Endo}},
  \bibinfo {author} {\bibfnamefont {A.}~\bibnamefont {Aspuru-Guzik}}, \bibinfo
  {author} {\bibfnamefont {S.~C.}\ \bibnamefont {Benjamin}}, \ and\ \bibinfo
  {author} {\bibfnamefont {X.}~\bibnamefont {Yuan}},\ }\href@noop {} {\bibfield
   {journal} {\bibinfo  {journal} {Reviews of Modern Physics}\ }\textbf
  {\bibinfo {volume} {92}},\ \bibinfo {pages} {015003} (\bibinfo {year}
  {2020})}\BibitemShut {NoStop}%
\bibitem [{\citenamefont {Bharti}\ \emph {et~al.}(2022)\citenamefont {Bharti},
  \citenamefont {Cervera-Lierta}, \citenamefont {Kyaw}, \citenamefont {Haug},
  \citenamefont {Alperin-Lea}, \citenamefont {Anand}, \citenamefont {Degroote},
  \citenamefont {Heimonen}, \citenamefont {Kottmann}, \citenamefont {Menke}
  \emph {et~al.}}]{bharti2022noisy}%
  \BibitemOpen
  \bibfield  {author} {\bibinfo {author} {\bibfnamefont {K.}~\bibnamefont
  {Bharti}}, \bibinfo {author} {\bibfnamefont {A.}~\bibnamefont
  {Cervera-Lierta}}, \bibinfo {author} {\bibfnamefont {T.~H.}\ \bibnamefont
  {Kyaw}}, \bibinfo {author} {\bibfnamefont {T.}~\bibnamefont {Haug}}, \bibinfo
  {author} {\bibfnamefont {S.}~\bibnamefont {Alperin-Lea}}, \bibinfo {author}
  {\bibfnamefont {A.}~\bibnamefont {Anand}}, \bibinfo {author} {\bibfnamefont
  {M.}~\bibnamefont {Degroote}}, \bibinfo {author} {\bibfnamefont
  {H.}~\bibnamefont {Heimonen}}, \bibinfo {author} {\bibfnamefont {J.~S.}\
  \bibnamefont {Kottmann}}, \bibinfo {author} {\bibfnamefont {T.}~\bibnamefont
  {Menke}},  \emph {et~al.},\ }\href@noop {} {\bibfield  {journal} {\bibinfo
  {journal} {Reviews of Modern Physics}\ }\textbf {\bibinfo {volume} {94}},\
  \bibinfo {pages} {015004} (\bibinfo {year} {2022})}\BibitemShut {NoStop}%
\bibitem [{\citenamefont {Kivlichan}\ \emph {et~al.}(2018)\citenamefont
  {Kivlichan}, \citenamefont {McClean}, \citenamefont {Wiebe}, \citenamefont
  {Gidney}, \citenamefont {Aspuru-Guzik}, \citenamefont {Chan},\ and\
  \citenamefont {Babbush}}]{kivlichan2018quantum}%
  \BibitemOpen
  \bibfield  {author} {\bibinfo {author} {\bibfnamefont {I.~D.}\ \bibnamefont
  {Kivlichan}}, \bibinfo {author} {\bibfnamefont {J.}~\bibnamefont {McClean}},
  \bibinfo {author} {\bibfnamefont {N.}~\bibnamefont {Wiebe}}, \bibinfo
  {author} {\bibfnamefont {C.}~\bibnamefont {Gidney}}, \bibinfo {author}
  {\bibfnamefont {A.}~\bibnamefont {Aspuru-Guzik}}, \bibinfo {author}
  {\bibfnamefont {G.~K.-L.}\ \bibnamefont {Chan}}, \ and\ \bibinfo {author}
  {\bibfnamefont {R.}~\bibnamefont {Babbush}},\ }\href@noop {} {\bibfield
  {journal} {\bibinfo  {journal} {Physical review letters}\ }\textbf {\bibinfo
  {volume} {120}},\ \bibinfo {pages} {110501} (\bibinfo {year}
  {2018})}\BibitemShut {NoStop}%
\bibitem [{\citenamefont {Babbush}\ \emph {et~al.}(2018)\citenamefont
  {Babbush}, \citenamefont {Wiebe}, \citenamefont {McClean}, \citenamefont
  {McClain}, \citenamefont {Neven},\ and\ \citenamefont
  {Chan}}]{low_depth_Chan}%
  \BibitemOpen
  \bibfield  {author} {\bibinfo {author} {\bibfnamefont {R.}~\bibnamefont
  {Babbush}}, \bibinfo {author} {\bibfnamefont {N.}~\bibnamefont {Wiebe}},
  \bibinfo {author} {\bibfnamefont {J.}~\bibnamefont {McClean}}, \bibinfo
  {author} {\bibfnamefont {J.}~\bibnamefont {McClain}}, \bibinfo {author}
  {\bibfnamefont {H.}~\bibnamefont {Neven}}, \ and\ \bibinfo {author}
  {\bibfnamefont {G.~K.-L.}\ \bibnamefont {Chan}},\ }\href {\doibase
  10.1103/PhysRevX.8.011044} {\bibfield  {journal} {\bibinfo  {journal} {Phys.
  Rev. X}\ }\textbf {\bibinfo {volume} {8}},\ \bibinfo {pages} {011044}
  (\bibinfo {year} {2018})}\BibitemShut {NoStop}%
\bibitem [{\citenamefont {Quantum}\ \emph {et~al.}(2020)\citenamefont
  {Quantum}, \citenamefont {Collaborators*†}, \citenamefont {Arute},
  \citenamefont {Arya}, \citenamefont {Babbush}, \citenamefont {Bacon},
  \citenamefont {Bardin}, \citenamefont {Barends}, \citenamefont {Boixo},
  \citenamefont {Broughton}, \citenamefont {Buckley} \emph
  {et~al.}}]{google2020hartree}%
  \BibitemOpen
  \bibfield  {author} {\bibinfo {author} {\bibfnamefont {G.~A.}\ \bibnamefont
  {Quantum}}, \bibinfo {author} {\bibnamefont {Collaborators*†}}, \bibinfo
  {author} {\bibfnamefont {F.}~\bibnamefont {Arute}}, \bibinfo {author}
  {\bibfnamefont {K.}~\bibnamefont {Arya}}, \bibinfo {author} {\bibfnamefont
  {R.}~\bibnamefont {Babbush}}, \bibinfo {author} {\bibfnamefont
  {D.}~\bibnamefont {Bacon}}, \bibinfo {author} {\bibfnamefont {J.~C.}\
  \bibnamefont {Bardin}}, \bibinfo {author} {\bibfnamefont {R.}~\bibnamefont
  {Barends}}, \bibinfo {author} {\bibfnamefont {S.}~\bibnamefont {Boixo}},
  \bibinfo {author} {\bibfnamefont {M.}~\bibnamefont {Broughton}}, \bibinfo
  {author} {\bibfnamefont {B.~B.}\ \bibnamefont {Buckley}},  \emph {et~al.},\
  }\href@noop {} {\bibfield  {journal} {\bibinfo  {journal} {Science}\ }\textbf
  {\bibinfo {volume} {369}},\ \bibinfo {pages} {1084} (\bibinfo {year}
  {2020})}\BibitemShut {NoStop}%
\bibitem [{\citenamefont {Geertsen}, \citenamefont {Rittby},\ and\
  \citenamefont {Bartlett}(1989)}]{bartlett89_57}%
  \BibitemOpen
  \bibfield  {author} {\bibinfo {author} {\bibfnamefont {J.}~\bibnamefont
  {Geertsen}}, \bibinfo {author} {\bibfnamefont {M.}~\bibnamefont {Rittby}}, \
  and\ \bibinfo {author} {\bibfnamefont {R.~J.}\ \bibnamefont {Bartlett}},\
  }\href {\doibase http://dx.doi.org/10.1016/0009-2614(89)85202-9} {\bibfield
  {journal} {\bibinfo  {journal} {Chem. Phys. Lett.}\ }\textbf {\bibinfo
  {volume} {164}},\ \bibinfo {pages} {57} (\bibinfo {year} {1989})}\BibitemShut
  {NoStop}%
\bibitem [{\citenamefont {Comeau}\ and\ \citenamefont
  {Bartlett}(1993)}]{bartlett93_414}%
  \BibitemOpen
  \bibfield  {author} {\bibinfo {author} {\bibfnamefont {D.~C.}\ \bibnamefont
  {Comeau}}\ and\ \bibinfo {author} {\bibfnamefont {R.~J.}\ \bibnamefont
  {Bartlett}},\ }\href {\doibase
  http://dx.doi.org/10.1016/0009-2614(93)89023-B} {\bibfield  {journal}
  {\bibinfo  {journal} {Chem. Phys. Lett.}\ }\textbf {\bibinfo {volume}
  {207}},\ \bibinfo {pages} {414} (\bibinfo {year} {1993})}\BibitemShut
  {NoStop}%
\bibitem [{\citenamefont {Stanton}\ and\ \citenamefont
  {Bartlett}(1993)}]{stanton93_5178}%
  \BibitemOpen
  \bibfield  {author} {\bibinfo {author} {\bibfnamefont {J.~F.}\ \bibnamefont
  {Stanton}}\ and\ \bibinfo {author} {\bibfnamefont {R.~J.}\ \bibnamefont
  {Bartlett}},\ }\href {\doibase http://dx.doi.org/10.1063/1.466019} {\bibfield
   {journal} {\bibinfo  {journal} {J. Chem. Phys.}\ }\textbf {\bibinfo {volume}
  {99}},\ \bibinfo {pages} {5178} (\bibinfo {year} {1993})}\BibitemShut
  {NoStop}%
\bibitem [{\citenamefont {Jeziorski}\ and\ \citenamefont
  {Paldus}(1989)}]{jeziorski1989valence}%
  \BibitemOpen
  \bibfield  {author} {\bibinfo {author} {\bibfnamefont {B.}~\bibnamefont
  {Jeziorski}}\ and\ \bibinfo {author} {\bibfnamefont {J.}~\bibnamefont
  {Paldus}},\ }\href@noop {} {\bibfield  {journal} {\bibinfo  {journal} {J.
  Chem. Phys.}\ }\textbf {\bibinfo {volume} {90}},\ \bibinfo {pages} {2714}
  (\bibinfo {year} {1989})}\BibitemShut {NoStop}%
\bibitem [{\citenamefont {Meissner}(1996)}]{meissner1996multiple}%
  \BibitemOpen
  \bibfield  {author} {\bibinfo {author} {\bibfnamefont {L.}~\bibnamefont
  {Meissner}},\ }\href@noop {} {\bibfield  {journal} {\bibinfo  {journal}
  {Chemical physics letters}\ }\textbf {\bibinfo {volume} {255}},\ \bibinfo
  {pages} {244} (\bibinfo {year} {1996})}\BibitemShut {NoStop}%
\bibitem [{\citenamefont {Meissner}(1998)}]{meissner1998fock}%
  \BibitemOpen
  \bibfield  {author} {\bibinfo {author} {\bibfnamefont {L.}~\bibnamefont
  {Meissner}},\ }\href@noop {} {\bibfield  {journal} {\bibinfo  {journal} {J.
  Chem. Phys.}\ }\textbf {\bibinfo {volume} {108}},\ \bibinfo {pages} {9227}
  (\bibinfo {year} {1998})}\BibitemShut {NoStop}%
\bibitem [{\citenamefont {Musial}\ and\ \citenamefont
  {Bartlett}(2008)}]{musial2008intermediate}%
  \BibitemOpen
  \bibfield  {author} {\bibinfo {author} {\bibfnamefont {M.}~\bibnamefont
  {Musial}}\ and\ \bibinfo {author} {\bibfnamefont {R.~J.}\ \bibnamefont
  {Bartlett}},\ }\href@noop {} {\bibfield  {journal} {\bibinfo  {journal} {The
  Journal of chemical physics}\ }\textbf {\bibinfo {volume} {129}},\ \bibinfo
  {pages} {044101} (\bibinfo {year} {2008})}\BibitemShut {NoStop}%
\bibitem [{\citenamefont {Meissner}(2022)}]{meissner2022new}%
  \BibitemOpen
  \bibfield  {author} {\bibinfo {author} {\bibfnamefont {L.}~\bibnamefont
  {Meissner}},\ }\href@noop {} {\bibfield  {journal} {\bibinfo  {journal}
  {Molecular Physics}\ ,\ \bibinfo {pages} {e2064355}} (\bibinfo {year}
  {2022})}\BibitemShut {NoStop}%
\bibitem [{\citenamefont {Low}\ and\ \citenamefont
  {Wiebe}(2018)}]{low2018hamiltonian}%
  \BibitemOpen
  \bibfield  {author} {\bibinfo {author} {\bibfnamefont {G.~H.}\ \bibnamefont
  {Low}}\ and\ \bibinfo {author} {\bibfnamefont {N.}~\bibnamefont {Wiebe}},\
  }\href@noop {} {\bibfield  {journal} {\bibinfo  {journal} {arXiv preprint
  arXiv:1805.00675}\ } (\bibinfo {year} {2018})}\BibitemShut {NoStop}%
\bibitem [{\citenamefont {Rajput}, \citenamefont {Roggero},\ and\ \citenamefont
  {Wiebe}(2021)}]{rajput2021hybridized}%
  \BibitemOpen
  \bibfield  {author} {\bibinfo {author} {\bibfnamefont {A.}~\bibnamefont
  {Rajput}}, \bibinfo {author} {\bibfnamefont {A.}~\bibnamefont {Roggero}}, \
  and\ \bibinfo {author} {\bibfnamefont {N.}~\bibnamefont {Wiebe}},\
  }\href@noop {} {\bibfield  {journal} {\bibinfo  {journal} {arXiv preprint
  arXiv:2109.03308}\ } (\bibinfo {year} {2021})}\BibitemShut {NoStop}%
\bibitem [{\citenamefont {Watkins}\ \emph {et~al.}(2022)\citenamefont
  {Watkins}, \citenamefont {Wiebe}, \citenamefont {Roggero},\ and\
  \citenamefont {Lee}}]{watkins2022time}%
  \BibitemOpen
  \bibfield  {author} {\bibinfo {author} {\bibfnamefont {J.}~\bibnamefont
  {Watkins}}, \bibinfo {author} {\bibfnamefont {N.}~\bibnamefont {Wiebe}},
  \bibinfo {author} {\bibfnamefont {A.}~\bibnamefont {Roggero}}, \ and\
  \bibinfo {author} {\bibfnamefont {D.}~\bibnamefont {Lee}},\ }\href@noop {}
  {\bibfield  {journal} {\bibinfo  {journal} {arXiv preprint arXiv:2203.11353}\
  } (\bibinfo {year} {2022})}\BibitemShut {NoStop}%
\bibitem [{\citenamefont {Jordan}\ and\ \citenamefont
  {Wigner}(1928)}]{Jordanwigner1928}%
  \BibitemOpen
  \bibfield  {author} {\bibinfo {author} {\bibfnamefont {P.}~\bibnamefont
  {Jordan}}\ and\ \bibinfo {author} {\bibfnamefont {E.}~\bibnamefont
  {Wigner}},\ }\href {\doibase 10.1007/BF01331938} {\bibfield  {journal}
  {\bibinfo  {journal} {Zeitschrift für Physik}\ }\textbf {\bibinfo {volume}
  {47}},\ \bibinfo {pages} {631} (\bibinfo {year} {1928})}\BibitemShut
  {NoStop}%
\bibitem [{\citenamefont {Lloyd}(1996)}]{lloyd1996universal}%
  \BibitemOpen
  \bibfield  {author} {\bibinfo {author} {\bibfnamefont {S.}~\bibnamefont
  {Lloyd}},\ }\href@noop {} {\bibfield  {journal} {\bibinfo  {journal}
  {Science}\ }\textbf {\bibinfo {volume} {273}},\ \bibinfo {pages} {1073}
  (\bibinfo {year} {1996})}\BibitemShut {NoStop}%
\bibitem [{\citenamefont {Farhi}, \citenamefont {Goldstone},\ and\
  \citenamefont {Gutmann}(2014)}]{farhi2014quantum}%
  \BibitemOpen
  \bibfield  {author} {\bibinfo {author} {\bibfnamefont {E.}~\bibnamefont
  {Farhi}}, \bibinfo {author} {\bibfnamefont {J.}~\bibnamefont {Goldstone}}, \
  and\ \bibinfo {author} {\bibfnamefont {S.}~\bibnamefont {Gutmann}},\
  }\href@noop {} {\bibfield  {journal} {\bibinfo  {journal} {arXiv preprint
  arXiv:1411.4028}\ } (\bibinfo {year} {2014})}\BibitemShut {NoStop}%
\bibitem [{\citenamefont {Farhi}\ \emph {et~al.}(2017)\citenamefont {Farhi},
  \citenamefont {Goldstone}, \citenamefont {Gutmann},\ and\ \citenamefont
  {Neven}}]{farhi2017quantum}%
  \BibitemOpen
  \bibfield  {author} {\bibinfo {author} {\bibfnamefont {E.}~\bibnamefont
  {Farhi}}, \bibinfo {author} {\bibfnamefont {J.}~\bibnamefont {Goldstone}},
  \bibinfo {author} {\bibfnamefont {S.}~\bibnamefont {Gutmann}}, \ and\
  \bibinfo {author} {\bibfnamefont {H.}~\bibnamefont {Neven}},\ }\href@noop {}
  {\bibfield  {journal} {\bibinfo  {journal} {arXiv preprint arXiv:1703.06199}\
  } (\bibinfo {year} {2017})}\BibitemShut {NoStop}%
\bibitem [{\citenamefont {Zhu}\ \emph {et~al.}(2020)\citenamefont {Zhu},
  \citenamefont {Tang}, \citenamefont {Barron}, \citenamefont
  {Calderon-Vargas}, \citenamefont {Mayhall}, \citenamefont {Barnes},\ and\
  \citenamefont {Economou}}]{zhu2020adaptive}%
  \BibitemOpen
  \bibfield  {author} {\bibinfo {author} {\bibfnamefont {L.}~\bibnamefont
  {Zhu}}, \bibinfo {author} {\bibfnamefont {H.~L.}\ \bibnamefont {Tang}},
  \bibinfo {author} {\bibfnamefont {G.~S.}\ \bibnamefont {Barron}}, \bibinfo
  {author} {\bibfnamefont {F.}~\bibnamefont {Calderon-Vargas}}, \bibinfo
  {author} {\bibfnamefont {N.~J.}\ \bibnamefont {Mayhall}}, \bibinfo {author}
  {\bibfnamefont {E.}~\bibnamefont {Barnes}}, \ and\ \bibinfo {author}
  {\bibfnamefont {S.~E.}\ \bibnamefont {Economou}},\ }\href@noop {} {\bibfield
  {journal} {\bibinfo  {journal} {arXiv preprint arXiv:2005.10258}\ } (\bibinfo
  {year} {2020})}\BibitemShut {NoStop}%
\bibitem [{\citenamefont {Kremenetski}\ \emph {et~al.}(2021)\citenamefont
  {Kremenetski}, \citenamefont {Hogg}, \citenamefont {Hadfield}, \citenamefont
  {Cotton},\ and\ \citenamefont {Tubman}}]{kremenetski2021quantum}%
  \BibitemOpen
  \bibfield  {author} {\bibinfo {author} {\bibfnamefont {V.}~\bibnamefont
  {Kremenetski}}, \bibinfo {author} {\bibfnamefont {T.}~\bibnamefont {Hogg}},
  \bibinfo {author} {\bibfnamefont {S.}~\bibnamefont {Hadfield}}, \bibinfo
  {author} {\bibfnamefont {S.~J.}\ \bibnamefont {Cotton}}, \ and\ \bibinfo
  {author} {\bibfnamefont {N.~M.}\ \bibnamefont {Tubman}},\ }\href@noop {}
  {\bibfield  {journal} {\bibinfo  {journal} {arXiv preprint arXiv:2108.13056}\
  } (\bibinfo {year} {2021})}\BibitemShut {NoStop}%
\bibitem [{\citenamefont {Herrman}\ \emph {et~al.}(2022)\citenamefont
  {Herrman}, \citenamefont {Lotshaw}, \citenamefont {Ostrowski}, \citenamefont
  {Humble},\ and\ \citenamefont {Siopsis}}]{herrman2022multi}%
  \BibitemOpen
  \bibfield  {author} {\bibinfo {author} {\bibfnamefont {R.}~\bibnamefont
  {Herrman}}, \bibinfo {author} {\bibfnamefont {P.~C.}\ \bibnamefont
  {Lotshaw}}, \bibinfo {author} {\bibfnamefont {J.}~\bibnamefont {Ostrowski}},
  \bibinfo {author} {\bibfnamefont {T.~S.}\ \bibnamefont {Humble}}, \ and\
  \bibinfo {author} {\bibfnamefont {G.}~\bibnamefont {Siopsis}},\ }\href@noop
  {} {\bibfield  {journal} {\bibinfo  {journal} {Scientific Reports}\ }\textbf
  {\bibinfo {volume} {12}},\ \bibinfo {pages} {1} (\bibinfo {year}
  {2022})}\BibitemShut {NoStop}%
\bibitem [{\citenamefont {Jankowski}\ and\ \citenamefont
  {Paldus}(1980)}]{jankowski1980applicability}%
  \BibitemOpen
  \bibfield  {author} {\bibinfo {author} {\bibfnamefont {K.}~\bibnamefont
  {Jankowski}}\ and\ \bibinfo {author} {\bibfnamefont {J.}~\bibnamefont
  {Paldus}},\ }\href@noop {} {\bibfield  {journal} {\bibinfo  {journal}
  {International Journal of Quantum Chemistry}\ }\textbf {\bibinfo {volume}
  {18}},\ \bibinfo {pages} {1243} (\bibinfo {year} {1980})}\BibitemShut
  {NoStop}%
\bibitem [{\citenamefont {Jankowski}\ and\ \citenamefont
  {Kowalski}(1996)}]{jankowski1996approximate}%
  \BibitemOpen
  \bibfield  {author} {\bibinfo {author} {\bibfnamefont {K.}~\bibnamefont
  {Jankowski}}\ and\ \bibinfo {author} {\bibfnamefont {K.}~\bibnamefont
  {Kowalski}},\ }\href@noop {} {\bibfield  {journal} {\bibinfo  {journal}
  {Chemical physics letters}\ }\textbf {\bibinfo {volume} {256}},\ \bibinfo
  {pages} {141} (\bibinfo {year} {1996})}\BibitemShut {NoStop}%
\bibitem [{\citenamefont {Jankowski}, \citenamefont {Grabowski},\ and\
  \citenamefont {Kowalski}(1998)}]{jankowski1998approximate}%
  \BibitemOpen
  \bibfield  {author} {\bibinfo {author} {\bibfnamefont {K.}~\bibnamefont
  {Jankowski}}, \bibinfo {author} {\bibfnamefont {I.}~\bibnamefont
  {Grabowski}}, \ and\ \bibinfo {author} {\bibfnamefont {K.}~\bibnamefont
  {Kowalski}},\ }\href@noop {} {\bibfield  {journal} {\bibinfo  {journal} {The
  Journal of chemical physics}\ }\textbf {\bibinfo {volume} {109}},\ \bibinfo
  {pages} {6255} (\bibinfo {year} {1998})}\BibitemShut {NoStop}%
\bibitem [{\citenamefont {Li}\ \emph {et~al.}(2000)\citenamefont {Li},
  \citenamefont {Grabowski}, \citenamefont {Jankowski},\ and\ \citenamefont
  {Paldus}}]{li2000approximate}%
  \BibitemOpen
  \bibfield  {author} {\bibinfo {author} {\bibfnamefont {X.}~\bibnamefont
  {Li}}, \bibinfo {author} {\bibfnamefont {I.}~\bibnamefont {Grabowski}},
  \bibinfo {author} {\bibfnamefont {K.}~\bibnamefont {Jankowski}}, \ and\
  \bibinfo {author} {\bibfnamefont {J.}~\bibnamefont {Paldus}},\ }in\
  \href@noop {} {\emph {\bibinfo {booktitle} {Advances in quantum
  chemistry}}},\ Vol.~\bibinfo {volume} {36}\ (\bibinfo  {publisher}
  {Elsevier},\ \bibinfo {year} {2000})\ pp.\ \bibinfo {pages}
  {231--251}\BibitemShut {NoStop}%
\end{thebibliography}

%\end{document}


\end{document}